\documentclass[12pt]{article}
\usepackage{psfig,epsf,rotating}

\begin{document}

%\compact
\baselineskip=15pt
\textwidth 15.0truecm
\textheight 21.0truecm
\topmargin 0.2in
\headsep 1.2cm

%\begin{center}

\title{Visible spectroscopic and photometric survey of Jupiter Trojans: final results on  dynamical families.
\footnote{Based on observations carried out at the European Southern
Observatory (ESO), La Silla, Chile, ESO proposals 71.C-0650, 73.C-0622, 74.C-0577} }

\author{Fornasier S.$^{1,2}$, Dotto E.$^3$, Hainaut O.$^4$, Marzari F.$^5$, \\
Boehnhardt H.$^6$, De Luise F.$^3$, Barucci M.A.$^2$}
%De Bergh C.$^6$

\maketitle
{\small
\noindent
$^1$ University of Paris 7, France \\
$^2$ LESIA -- Paris Observatory, France.\\
$^3$ INAF -- Osservatorio Astronomico di Roma, Italy; \\
$^4$ European Southern Observatory, Chile;\\
$^5$ Dipartimento di Fisica, Universit\`a di Padova, Italy; \\
$^6$ Max-Planck Institute for Solar System Research, Katlenburg-Lindau, Germany\\
}

\maketitle

\noindent
Submitted to Icarus: December 2006\\
e-mail: sonia.fornasier@obspm.fr\\
fax: +33145077144, phone: +33145077746\\
%Manuscript pages: 69; Figures: 14; Tables: 8\\

\vspace{3cm}

{\bf Running head}: Investigation of Dynamical Families of Jupiter Trojans 

\vspace{1cm}

\noindent

{\it Send correspondence to:}\\
Sonia Fornasier  \\
LESIA-Observatoire de Paris  \\
Batiment 17 \\
5, Place Jules Janssen \\
92195 Meudon Cedex \\
France\\
e-mail: sonia.fornasier@obspm.fr\\
fax: +33145077144\\
phone: +33145077746\\

\newpage
\vspace{2.5cm}

\begin{abstract}

We present the results of a visible spectroscopic and photometric 
survey of Jupiter Trojans belonging to different dynamical
families. The survey was carried out at the 3.5m New Technology Telescope (NTT) of
the European Southern Observatory (La Silla, Chile) in April 2003, May
2004 and January 2005.  We obtained data on 47 objects, 23 belonging
to the L5 swarm and 24 to the L4 one.  These data together with those
already published by Fornasier et al. (2004a) and Dotto et al. (2006),
acquired since November 2002, constitute a
total sample of visible spectra for 80 objects.\\ 
The survey allows us to investigate six families (Aneas, Anchises, Misenus, Phereclos, Sarpedon, Panthoos) in the L5 cloud and four L4 families (Eurybates, Menelaus, 1986 WD and 1986 TS6).  
The sample that we measured is dominated by D--type asteroids, with the exception of the Eurybates family in the L4 swarm, where there is a dominance of C-- and P--type asteroids. \\
All the spectra that we obtained are featureless with
the exception of some Eurybates members, where a drop--off of the reflectance is detected shortward of 5200 \AA. Similar features are seen in main belt C--type asteroids and commonly attributed to the intervalence
charge transfer transition in oxidized iron.  \\ 
Our sample comprises fainter and smaller Trojans as compared to the literature's data and allows us to investigate the properties of objects with estimated diameter smaller than 40--50 km. The analysis of the spectral slopes and colors versus the estimated diameters shows that the blue and red objects have indistinguishable size distribution, so any relationship between size and spectral slopes has been found. \\ 
To fully investigate the Trojans population, we include in our analysis 62 spectra of Trojans available in literature, resulting in a total sample of 142 objects. Although the mean spectral behavior of L4 and L5
Trojans is indistinguishable within the uncertainties, we find that
the L4 population is more heterogeneous and that it has a higher abundance of bluish objects as compared to the L5 swarm. \\ 
Finally, we perform a statistical investigation of the Trojans's spectra property distributions as a function of their orbital and physical parameters, and in comparison with other classes of minor bodies in the outer Solar System. 
Trojans at lower inclination appear significantly bluer than those at higher inclination, but this effect is strongly driven by the Eurybates family.
The mean colors of the Trojans are similar to those of short period comets and neutral Centaurs, but their color distributions are different. 
\end{abstract}

Keywords:  Trojan Asteroids -- Photometry -- Spectroscopy --
-- Asteroids families

\newpage

%==============================================================================
%
			\section{Introduction}
%
%==============================================================================

Jupiter Trojans are small bodies of the Solar System located in the
Jupiter Lagrangian points L4 and L5. Up to now more than 2000 Trojans have been discovered, $\sim$ 1150 belonging
to the L4 cloud and $\sim$ 950 to the L5 one. The number of L4 Trojans
with radius greater than 1 km is estimated to be around 1.6
$\times$10$^{5}$ (Jewitt et al., 2000), comparable with the estimated
main belt population of similar size. \\
The debate about the origin of Jupiter Trojans and how they were
trapped in librating orbits around the Lagrangian points is still open
to several possibilities.  
Considering that Trojans have orbits stable over the age of the Solar
System (Levison et al, 1997, Marzari et al. 2003) their origin must date
back to the early phase of the solar system formation. Some authors
(Marzari \& Scholl, 1998a,b; Marzari et al., 2002) suggested that they formed 
very close to their present location and were trapped during the growth of Jupiter.
Morbidelli et al. (2005) suggested that Trojans formed in the Kuiper belt and were
subsequently captured in the Jupiter L4 and L5 Lagrangian points
during planetary migration, just after Jupiter and Saturn crossed
their mutual 1:2 resonances. In this scenario, Jupiter Troians would give important clues on the
composition and accretion of bodies in the outer regions of the solar
nebula.

Several theoretical studies conclude that Jupiter Trojan clouds are at
least as collisionally evolved as main belt asteroids (Shoemaker et
al., 1989; Binzel \& Sauter, 1992; Marzari et al., 1997; Dell'Oro et al., 1998). This result
is supported by the identification of several dynamical families, both
in the L4 and L5 swarm (Shoemaker et al., 1989, Milani, 1993, Beaug\'e
and Roig, 2001).  \\
Whatever the Trojan origin is, it is plausible to assume that they
formed beyond the frost line and that they are primitive bodies, are possibly composed of anhydrous silicates and organic compounds, and possibly still contain ices in their interior.
Several observations of Trojans in the near infrared region (0.8-2.5
$\mu$m) have failed to clearly detect any absorption features indicative of water ice (Barucci et al, 1994; Dumas et al, 1998; Emery \& Brown, 2003, 2004; Dotto et al., 2006). Also in the visible range Trojan
spectra appear featureless (Jewitt \& Luu, 1990; Fornasier et al.,
2004a, Bendjoya et al., 2004; Dotto et al., 2006). Up to now only 2
objects (1988 BY1 and 1870 Glaukos) show the possible presence of
faint bands (Jewitt \& Luu, 1990). However, these bands are comparable to the peak to peak noise and are not yet confirmed. \\ 
Recently, mineralogical features have been detected in emissivity spectra of three Trojan asteroids measured by the Spitzer Space Telescope. These features are interpreted as indicating the presence of fine-grained silicates on the surfaces (Emery et al. 2006).

Several questions about Jupiter Trojans' dynamical origin, physical
properties, composition and link with other groups of minor bodies
such as outer main belt asteroids, cometary nuclei, Centaurs and KBOs 
are still open.\\
In order to shed some light on these questions, we have carried out a spectroscopic and photometric survey of Jupiter
Trojans at the 3.5m New Technology Telescope (NTT) of the European Southern Observatory (La Silla, Chile) and at the 3.5m Telescopio
Nazionale Galileo (TNG), La Palma, Spain. In this paper we present new
visible spectroscopic and photometric data, obtained during 7 observing nights, carried out at ESO-NTT on April 2003, May 2004, and January 2005, for a total of 47 objects belonging to the
L5 (23 objects) and L4 (24 objects) swarms.  Considering also the
results already published in Fornasier et al. (2004a) and Dotto et
al. (2006), obtained in the framework of the same project, we
collected a total sample of 80 Jupiter Trojan visible spectra, 47
belonging to the L5 clouds and 33 to the L4.  This is the largest
homogeneous data set available up to now on these primitive
asteroids. 

The principal aim of our survey was the investigation of
Jupiter Trojans belonging to different dynamical families.  In fact,
since dynamical families are supposed to be formed from the
collisional disruption of parent bodies, the investigation of the
surface properties of small and large family members can help in understanding the nature of these dynamical groups and might
provide a glimpse of the interior structure of the larger primordial
parent bodies. \\
We also present an analysis of the visible spectral
slopes for all the data in our survey along with those available in the literature, for a total sample of 142 Trojans. \\
This enlarged sample allowed us to carry out a significant statistical investigation of the Trojans' spectral property distributions, as a function of their
orbital and physical parameters, and in comparison with other classes of minor
bodies in the outer Solar System. We also discuss the spectral slope
distribution within the Trojan families.

%==============================================================================
%
	      \section{Observations and data reduction}
%
%==============================================================================

[HERE TABLE 1 AND 2]

The data were obtained in the visible range during 3 different
observing runs at ESO-NTT: 10 and 11 April 2003 for the
spectroscopic and photometric investigation of 6 members of the 4035
1986~WD and 1 member of 1986~TS6 families; 25 and 26 May 2004 for a
spectroscopic survey of L4 Eurybates family; 17, 18, and 19 January
2005 for the spectroscopic and photometric investigation of 5
Anchises, 6 Misenus, 5 Panthoos, 2 Cloanthus, 2 Sarpedon and 3
Phereclos family members (L5 swarm).

We selected our targets from the list of Jupiter Trojan families provided by Beaug\'e and Roig (2001 and P.E.Tr.A. Project at www.daf.on.br/~froig/petra/). \\
The authors have used a cluster-detection algorithm called Hierarchical
Clustering Method (HCM, e.g. Zappal\`a et al., 1990) to find asteroid families among Jupiter Trojans starting from a
data--base of semi-analytical proper elements (Beaug\'e \& Roig, 2001). The identification of families is performed by
comparing the mutual distances with a suitable metric in the proper
elements' space. The clustering chain is halted when the mutual distance,
measuring the incremental velocity needed for orbital change after the
putative parent body breakup, is larger than a fixed cut-off value. A
lower cutoff implies a higher statistical significance of the family.
Since families in L4 are on average more robust than those around L5
(Beaug\'e and Roig, 2001), we prefer to adopt a cutoff of 100 $m/s$ for the L4 cloud and of 150 $m/s$
for L5. For the very robust Eurybates family we decided to limit our
survey to those family members defined with a cutoff of 70 $m/s$.

All the data were acquired using the EMMI instrument, equipped with a 2x1 mosaic of 2048$\times$4096 MIT/LL CCD with square 15$\mu$m pixels.
For the spectroscopic investigation during May 2004 and January 2005 runs
we used the grism \#1 (150 gr/mm) in RILD mode to cover the wavelength 
range 4100--9400 \AA\ with a dispersion of 3.1 \AA/px (200
\AA/mm) at the first order, while on April 2003 we used a different
grism, the \#7 (150 gr/mm), covering the spectral range 5200--9500
\AA, with a dispersion of 3.6 \AA/px at the first order.  April 2003
and January 2005 spectra were taken through a 1 arcsec wide slit,
while during May 2004 we used a larger slit (1.5 arcsec).  The slit
was oriented along the parallactic angle during all the observing runs
in order to avoid flux loss due to the atmospheric differential
refraction. \\ 
For most objects, the total exposure time was divided into several (usually 2-4) shorter acquisitions. This allowed us to check the asteroid position in the slit before each acquisition, and correct the telescope
pointing and/or tracking rates if necessary. During each night we also recorded bias, flat--field, calibration lamp
(He-Ar) and several (6-7) spectra of solar analog stars measured at different airmasses, covering the airmass range of the science targets. During 17 January 2005, part of the night was
lost due to some technical problems and only 2 solar analog stars were
acquired.  The ratio of these 2 stars show minimal variations (less
than 1\%) in the 5000--8400 \AA\ range, but higher differences at the
edges of this range. For this reason we omit the spectral
region below 4800 \AA\ for most of the asteroids acquired that
night.  

\begin{figure}
\centerline{\psfig{file=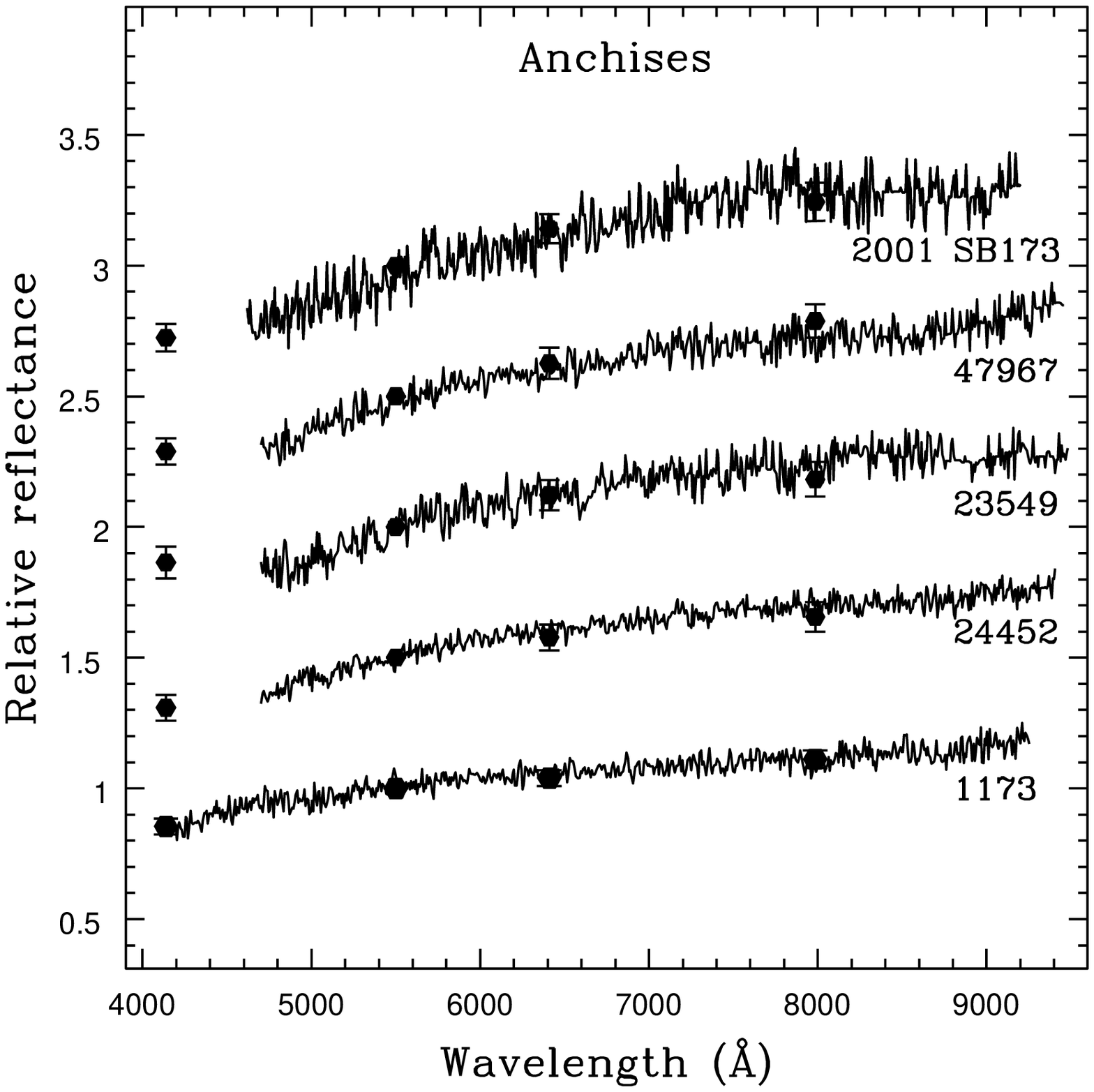,width=15truecm,angle=0}}
\caption{}
\label{fig1}
\end{figure}

\begin{figure}
\centerline{\psfig{file=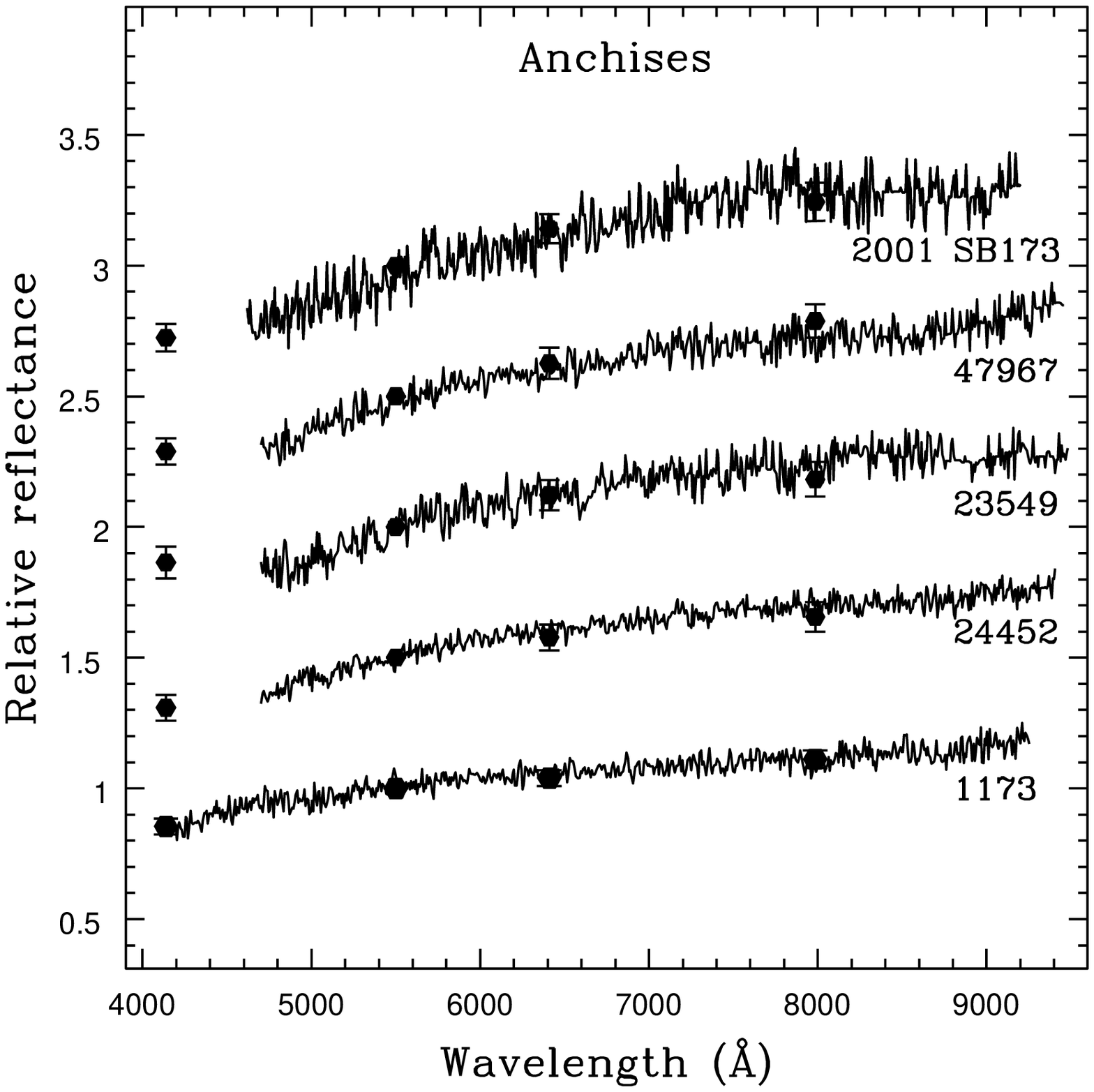,width=15truecm,angle=0}}
\caption{}
\label{fig2}
\end{figure}

\begin{figure}
\centerline{\psfig{file=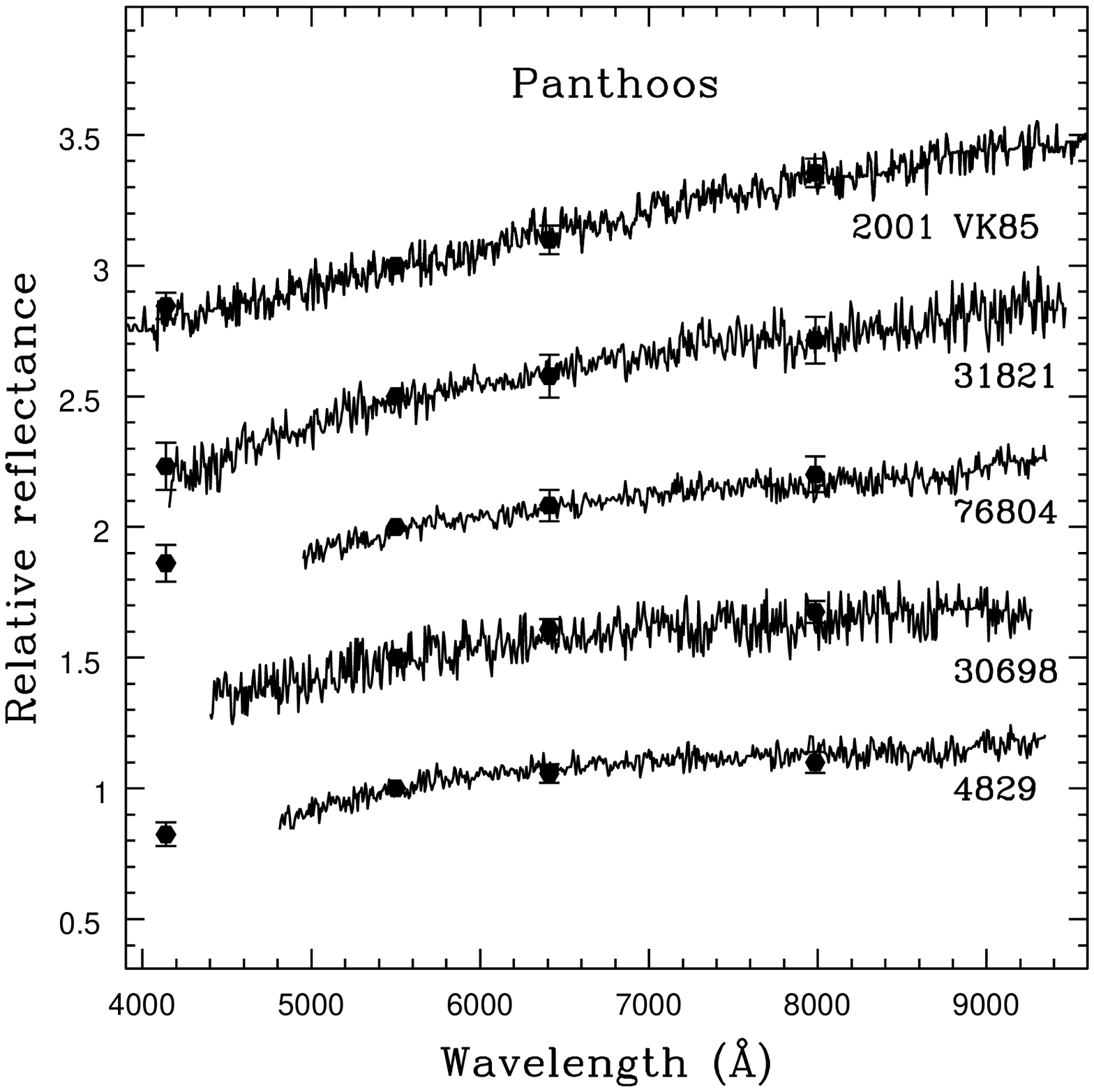,width=15truecm,angle=0}}
\caption{}
\label{fig3}
\end{figure}

\begin{figure}
\centerline{\psfig{file=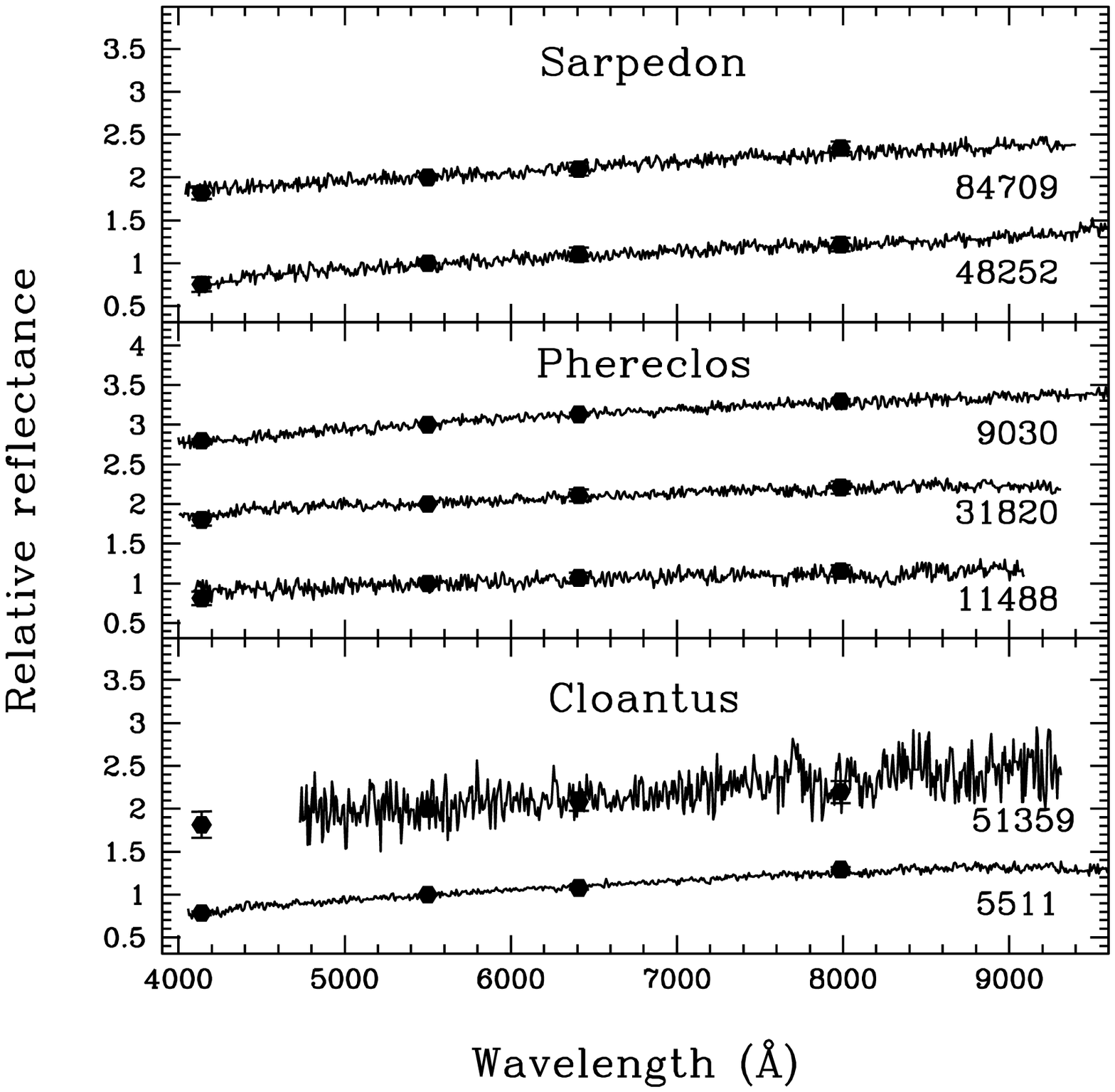,width=15truecm,angle=0}}
\caption{}
\label{fig4}
\end{figure}

\begin{figure}
\centerline{\psfig{file=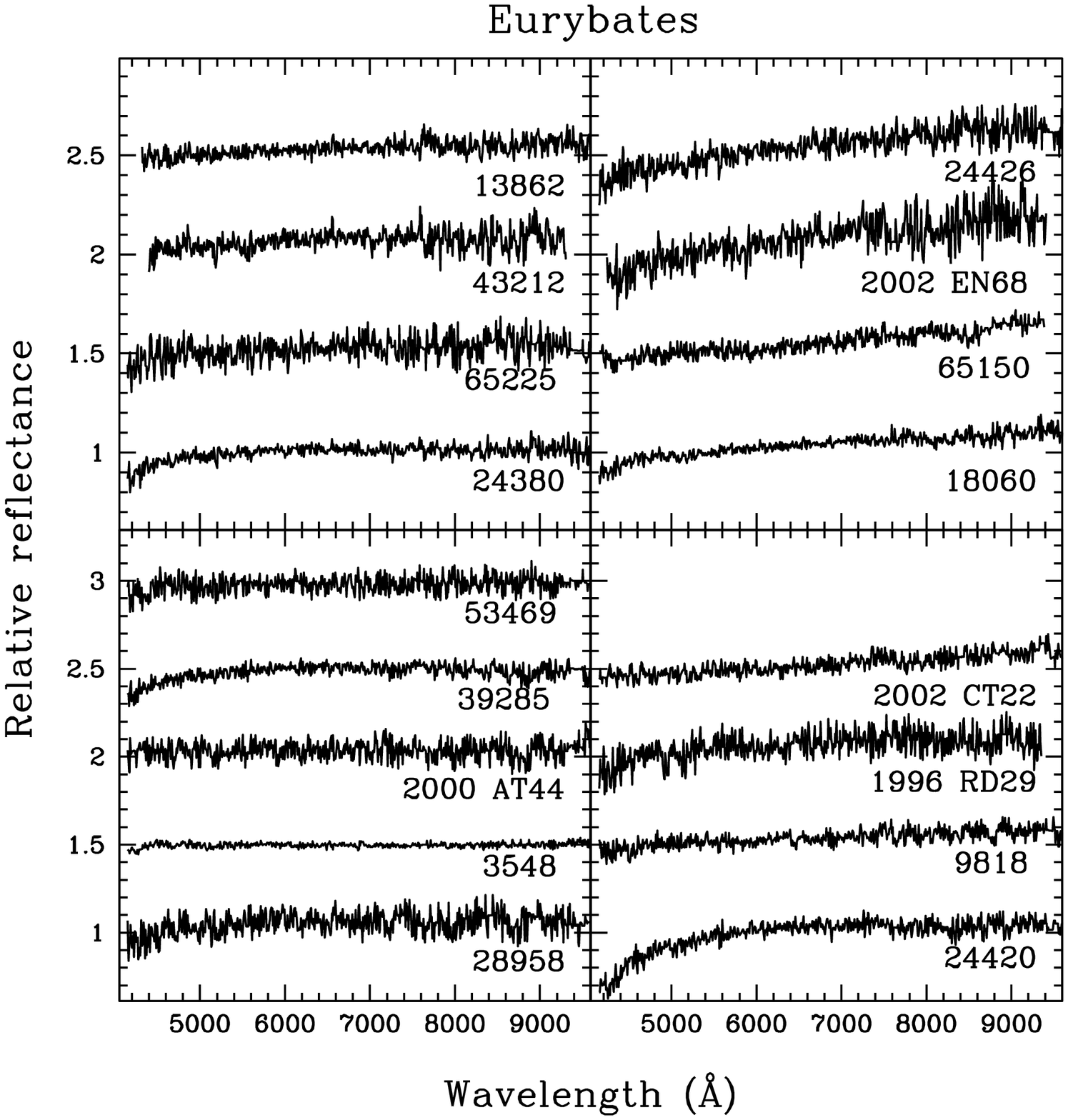,width=15truecm,angle=0}}
\caption{}
\label{fig5}
\end{figure}

The spectra were reduced using ordinary procedures of data
reduction as described in Fornasier et al. (2004a).  The
reflectivity of each asteroid was obtained by dividing its spectrum
by that of the solar analog star closest in time and airmass to
the object. Spectra were finally smoothed with a median filter 
technique, using a box of 19 pixels in the spectral direction for each point of the 
spectrum. The threshold was set to 0.1, meaning that the original value was replaced by 
the median value if the median value differs by more than 10\% from the original one.
The obtained spectra are shown in Figs.~\ref{fig1}--\ref{fig5}.
In Table~\ref{spec_obsL5} and Table~\ref{spec_obsL4} we report the 
circumstances of the observations and the solar analog stars used 
respectively for the  L5 and L4 family members.  

[TABLE 3]

The broadband color data were obtained during the April 2003 and
January 2005 runs just before the Trojans' spectral observation. We
used the RILD mode of EMMI for wide field imaging with the Bessell-type B, V, R, and I filters (centered respectively at 4139, 5426, 6410 and
7985\AA).  The observations were carried out in a $2\times 2$ binning mode,
yielding a pixel scale of 0.33 arcsec/pixel.  The exposure time varied
with the object magnitude: typically it was about 12-90s in V, 30-180s
in B, 12-70s in R and I filters.  \\ 
The CCD images were reduced and calibrated with a standard method
(Fornasier et al., 2004a), and absolute calibration was obtained through the observations of several Landolt fields (Landolt, 1992).
The instrumental magnitudes were measured using aperture photometry with
an integrating radius typically about three times the average seeing,
and sky subtraction was performed using a 5-10 pixels wide annulus
around each object. \\ 
The results are reported in Table~\ref{photometry}. From the visual inspection and the radial profiles analysis of the images, no coma was detected for any of the observed Trojans. \\ 
On May 2004, as the sky conditions were clear but not
photometric, we did not perform photometry of the Eurybates family
targets.\\

%==============================================================================
% 
			  \section{Results}
%
%==============================================================================

[TABLE 4 AND 5]

For each Trojan we computed the slope $S$ of the spectral continuum
using a standard least squared technique for a linear fit in the
wavelength range between 5500 and 8000 \AA. The choice of these
wavelength limits has been driven by the spectral coverage of our
data. We choose 5500 \AA\ as the lower limit because of the
different instrumental setup used during different observing runs
(with some spectra starting at wavelength $\ge$ 5200 \AA), while
beyond 8000 \AA\ our spectra are generally noisier due to a combination
of the CCD drop-off in sensitivity and the presence of the strong
atmospheric water bands.  \\ 
The computed slopes and errors are listed in Table~\ref{spec_L5}
and~\ref{spec_L4}.  The reported error bars take into account the
$1\sigma$ uncertainty of the linear fit plus 0.5\%/$10^3$\AA\
attributable to the use of different instruments and solar analog
stars (estimated from the different efficiency of the grism used, and from flux losses due to different slit apertures). In Table~\ref{spec_L5} and~\ref{spec_L4} we also report the
taxonomic class derived following the Dahlgren \& Lagerkvist (1995)
classification scheme. 

In the L5 cloud we find 27 D--, 3 DP--, 2 PD--, and 1 P--type
objects.  In the L4 cloud we find
 10 C--type and 7 P--type objects
inside the Eurybates family, while for the Menelaus, 1986~TS6 and
1986~WD families, including the data published in Dotto et
al. (2006), we get 9 D--, 3 P--, 3C--, and 1 DP--type asteroids.

The majority of the spectra are featureless, although some of the
observed Eurybates' members show weak spectral absorption features
(Fig.~\ref{fig5}). These features are discussed in the following section.

We derived an estimated absolute magnitude H by scaling the measured V magnitude to 
$r = \Delta =$ 1 AU and to zero phase assuming G=0.15 (Bowell et al., 1989). The estimated H magnitude of each Trojan might be skewed uncertain rotational phase, as the lightcurve amplitudes of Trojans might vary up to 1 magnitude. 
In order to investigate possible size dependence
inside each family, and considering that IRAS diameters are available
for very few objects, we estimate the size using the following relationship:\\
\[ D= \frac{1329 \times 10^{-H/5}}{\sqrt{p}} \] 
where D is the asteroid diameter, p is the geometric albedo, and H is
the absolute magnitude. We use H derived from our observations when available, and from the ASTORB.DAT file
(Lowell observatory) for the Eurybates members, for which we did not
carry out visible photometry.  We evaluated the diameter for an albedo
range of 0.03--0.07, assuming a mean albedo of 0.04 for these dark
asteroids (Fernandez et al., 2003). The resulting
D values are reported in Tables~\ref{spec_L5} and~\ref{spec_L4}.

\subsection{Dynamical families: L5 swarm}

\subsubsection{Anchises}\label{sect:anchises}

[FIGURE 1]

We investigated 5 of the 15 members of the Anchises family
(Fig.~\ref{fig1}): 1173 Anchises, 23549 1994 ES6, 24452 2000 QU167,
47967 2000 SL298 and 124729 2001 SB173 on 17 January 2005.
For 4 out of 5 observed objects we omit the spectral range below 4800\AA\ due to low S/N ratio and problems with the solar analog stars. The spectral behavior is confirmed by photometric data
(see Table~\ref{photometry}).  All the obtained spectra are
featureless. 

The Anchises family survives at a cutoff corresponding to relative
velocities of 150 m/s.  The biggest member, 1173 Anchises, has a
diameter of 126 km (IRAS data) and has the lowest spectral slope (3.9
\%/$10^{3}$\AA) among the investigated family members. It is
classified as P--type, while the other 4 members are all D--types. Anchises was previously observed in the 4000-7400\AA\ region by Jewitt \& Luu
(1990), who reported a spectral slope of 3.8 \%/$10^{3}$\AA, in
perfect agreement with the value we found.  The three 19-29 km sized
objects have a steeper spectral slope (7.4-9.2 \%/$10^{3}$\AA), while
the smallest object, 2001 SB173 (spectral slope = 14.78$\pm$0.99 \%/$10^{3}$\AA) is the reddest one
(Table~\ref{spec_L5}). 

Even with the uncertainties in the albedo and diameter, a
slope--size relationship is evident among the observed objects, with
smaller--fainter members redder than larger ones (Fig.~\ref{fig7}). 

\subsubsection{Misenus}\label{sect:misenus}

[FIGURE 2]

For this family we investigated 6 members (11663 1997 GO24, 32794 1989
UE5, 56968 2000 SA92, 99328 2001 UY123, 105685 2000 SC51 and 120453
1988 RE12) out of the 12 grouped at a relative velocity of 150
m/s. The family survives with the same members also at a stringent cut-off velocity of 120 m/s. The spectra, together with magnitude color indices transformed
into linear reflectance, are shown in Fig.~\ref{fig2}, while the color
indices are reported in Table~\ref{photometry}.  All the spectra are
featureless with different spectral slope values covering the 4.6--15.9 \%/$10^{3}$\AA\ range (Table~\ref{spec_L5}): 1988 RE12 has the lowest spectral 
slope and is classified as P--type, 3 objects (11663, 32794 and 2000 SC51) are in
the transition region between P-- and D-- type, with very similar spectral
behavior, while the two other observed members are D--types.  Of these
last, 56968 has the highest spectral slope not only inside the family
(15.86 \%/$10^{3}$\AA) but also inside the whole L5 sample analyzed in this paper. 

All the investigated Misenus members are quite faint and have diameters of a few tens of kilometers. No clear size-slope
relationship has been found inside this family (Fig.~\ref{fig7}). \\
No other data on the Misenus family members are available in the
literature, so we do not know if the large gap between the spectral
slope of 56968 and those of the other 5 investigated objects is real
or it could be filled by other members not yet observed. If real, 56968 can be an interloper inside the family.\\

\subsubsection{Panthoos}\label{sect:panthoos}

[FIGURE 3]

The Panthoos family has 59 members for a relative velocity cutoff of 150 m/s.
We obtained new spectroscopic and photometric data of 5 members: 4829 Sergestus, 30698 Hippokoon, 31821
1999 RK225, 76804 2000 QE and 111113 2001 VK85 (Fig.~\ref{fig3}).
Three objects presented by Fornasier et al. (2004a) as belonging to the
Astyanax family (23694 1997 KZ3, 32430 2000 RQ83, 30698 Hippokoon) and
one to the background population (24444 2000 OP32) are now included
among the members of the Panthoos family. Periodic updates
of the proper elements can change the family membership.  In particular
the Astyanax group disappeared in the latest revision of dynamical
families, and its members are now in the Panthoos family within a
cutoff of 150m/s. The Panthos family survives also a cutoff of 120 m/s, with 7 members, and 90 m/s, with 6 members. \\
 We observed 30698 Hippokoon during two different
runs (on 9 Nov. 2002 and on 18 Jan. 2005), and both spectral slopes
and colors are in agreement inside the error bars (see
 Table~\ref{photometry}, Table~\ref{spec_L5}, and Fornasier et al., 2004a).  No other data on the Panthoos family are available in the
literature. 

The analysis of the 8 members (for 24444 only
photometry is available) show featureless spectra with slopes that
seem to slightly increase as the asteroid size decreases
(Table~\ref{spec_L5} and Fig.~\ref{fig7}). However, all the members have
dimensions very similar within the uncertainties, making it difficult
for any slope-size relationship to be studied. The largest member, 4829 Sergestus, is a PD--type with a slope of about 5
\%/$10^{3}$\AA, while all the other investigated members are D--types.

\subsubsection{Cloantus}\label{sect:cloantus}

[FIGURE 4]

We observed only 2 out of 8 members of the Cloantus family (5511 Cloanthus 
and 51359 2000 SC17, see Fig.~\ref{fig4}) as grouped at a cutoff corresponding
to relative velocities of 150 m/s. This family survives at a stringent cutoff and 3 members (including the two that we observed) also survive 
for relative velocities of 60 m/s.
Both of the observed objects are D--types with very similar, featureless, 
reddish spectra 
(Table~\ref{spec_L5} and Fig.~\ref{fig7}). 5511 Cloanthus was observed also by Bendjoya et al. 
(2004), who found a slope of 13.0$\pm$0.1 \%/$10^{3}$\AA\ in the 5000-7500 \AA\ wavelength 
range, while we measure a value of 10.84$\pm$0.15  \%/$10^{3}$\AA. 
Our spectrum has a higher S/N ratio than the spectrum by Bendjoya et al. (2004), and it is perfectly matched by our measured color indices that confirm the spectral slope.
This difference cannot be caused by the slightly different spectral ranges used to measure the slope, but could possibly be due to heterogeneous surface composition.

\subsubsection{Phereclos}\label{sect:phereclos}

The Phereclos family comprises 15 members at a cutoff of 150 m/s. The family survives with 8 members also at a cutoff of 120m/s.  We
obtained spectroscopic and photometric data of 3 members (9030 1989
UX5, 11488 1988 RM11 and 31820 1999 RT186, see Fig~\ref{fig4}), that,
together with the 4 spectra (2357 Phereclos, 6998 Tithonus, 9430 1996
HU10, 18940 2000QV49) already presented by Fornasier et al. (2004a),
allow us to investigate about half of the Phereclos family population defined at a cutoff of 150m/s.
The spectral slope of these objects, all classified as D--type except one
PD--type (11488), varies from 5.3 to 11.3 \%/$10^{3}$\AA\
(Table~\ref{spec_L5}).  The size of the family members ranges from
about 20 km in diameter for 31820 to 95 km for 2357, but we do not
observe any clear slope-diameter relationship (Fig.~\ref{fig7} and
Table~\ref{spec_L5}).

\subsubsection{Sarpedon}\label{sect:sarpedon}

We obtained new spectroscopic and photometric data of 2 members of the
Sarpedon family (48252 2001 TL212 and 84709 2002 VW120), whose spectra
and magnitude color indices are reported in Fig.~\ref{fig4} and
Table~\ref{spec_L5}. Including the previous observations (Fornasier
et al., 2004a) of 4 other members (2223 Sarpedon, 5130 Ilioneus, 17416
1988 RR10, and 25347 1999 RQ116), we have measurements of 6 of the 21 members of this
family dynamically defined at a cutoff of 150 m/s.  All the 6 aforementioned objects, except 25347, constitute a robust clustering which
survives up to 90 m/s with 9 members. The cluster which contains (2223) Sarpedon was
also recognized as a family by Milani (1993). 

All the 6 investigated members have very similar colors
(see Table~\ref{photometry}) and spectral behavior. The spectral slope
(Fig.~\ref{fig7}) varies over a very restricted range, from 9.6 to
11.6 \%/$10^{3}$\AA\ (Table~\ref{spec_L5}), despite a significant
variation of the estimated size (from the 18 km of 17416 to the 105 km
of 2223). Consequently, the surface composition of the Sarpedon family
members appears to be very homogeneous.

% 
% % 
% % 
%   L4 
% % 
% %

\subsection{Dynamical families: L4 swarm}
\subsubsection{Eurybates}\label{sect:eurybates}

[FIGURE 5]

Eurybates family members were observed in May 2004. The selection
of the targets was made on the basis of a very stringent cutoff,
corresponding to relative velocities of 70 m/s, that gives a family
population of 28 objects. We observed 17 of these members (see
Table~\ref{spec_obsL4}) that constitute a very robust clustering in
the space of the proper elements: all the members we studied, except
2002 CT22, survive at a cutoff of 40 m/s.  

The spectral behavior of these objects (Fig.~\ref{fig5}) is quite
homogeneous with 10 asteroids classified as C--type and 7 as
P--type. The spectral slopes (Table~\ref{spec_L4}) range from neutral
to moderately red (from -0.5 to 4.6 \%/$10^{3}$\AA). 
The slopes of six members are close to zero (3 slightly negative) with solar-like colors. The asteroids 18060, 24380, 24420, and 39285, all classified as C--types, 
clearly show a drop off of reflectance for wavelength shorter
than 5000--5200 \AA. The presence of the same feature in the spectra
of 2 other members (1996 RD29 and 28958) is less certain due to
the lower S/N ratio. This absorption is commonly seen on main belt C--type asteroids (Vilas 1994; Fornasier et al. 1999), where is due to the intervalence charge
transfer transitions (IVCT) in oxidized iron, and is often coupled with
other visible absorption features related to the presence of aqueous
alteration products (e.g. phyllosilicates, oxides, etc).  These IVCTs
comprise multiple absorptions that are not uniquely indicative of
phyllosilicates, but are present in the spectrum of any object
containing Fe$^{2+}$ and Fe$^{3+}$ in its surface material (Vilas
1994).  Since no other phyllosilicate absorption features are present
in the C-type spectra of the Eurybates family, there is no evidence
that aqueous alteration processes occurred on the surface of these
bodies. 

In Fig.~\ref{fig8} we show the spectral slopes versus the estimated
diameters for the Eurybates family members.  All the observed objects,
except the largest member (3548) that has a diameter of about 70 km and exhibit a neutral ($\sim$ solar-like) spectral slope, are smaller
than $\sim$ 40 km and present both neutral and moderately red colors.
The spectral slopes are strongly clustered around $S=2 \% / 10^3$\AA,
with higher $S$ values restricted to smaller objects (D$<$ 25 km).
  
\subsubsection{1986~WD}\label{sect:1986wd}

[FIGURE 6]

\begin{figure}
\centerline{\psfig{file=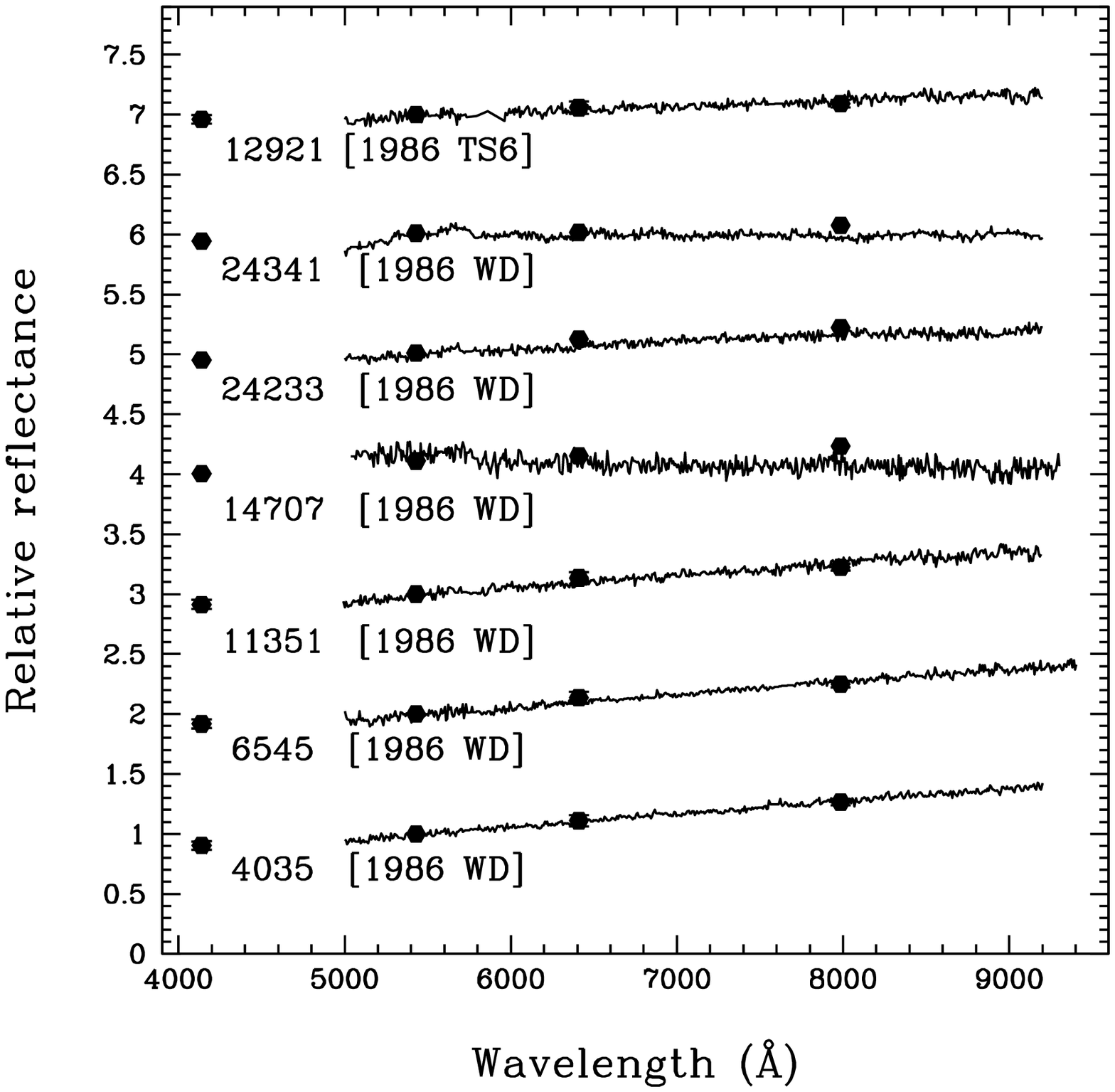,width=15truecm,angle=0}}
\caption{}
\label{fig6}
\end{figure}

\begin{figure}
\centerline{\psfig{file=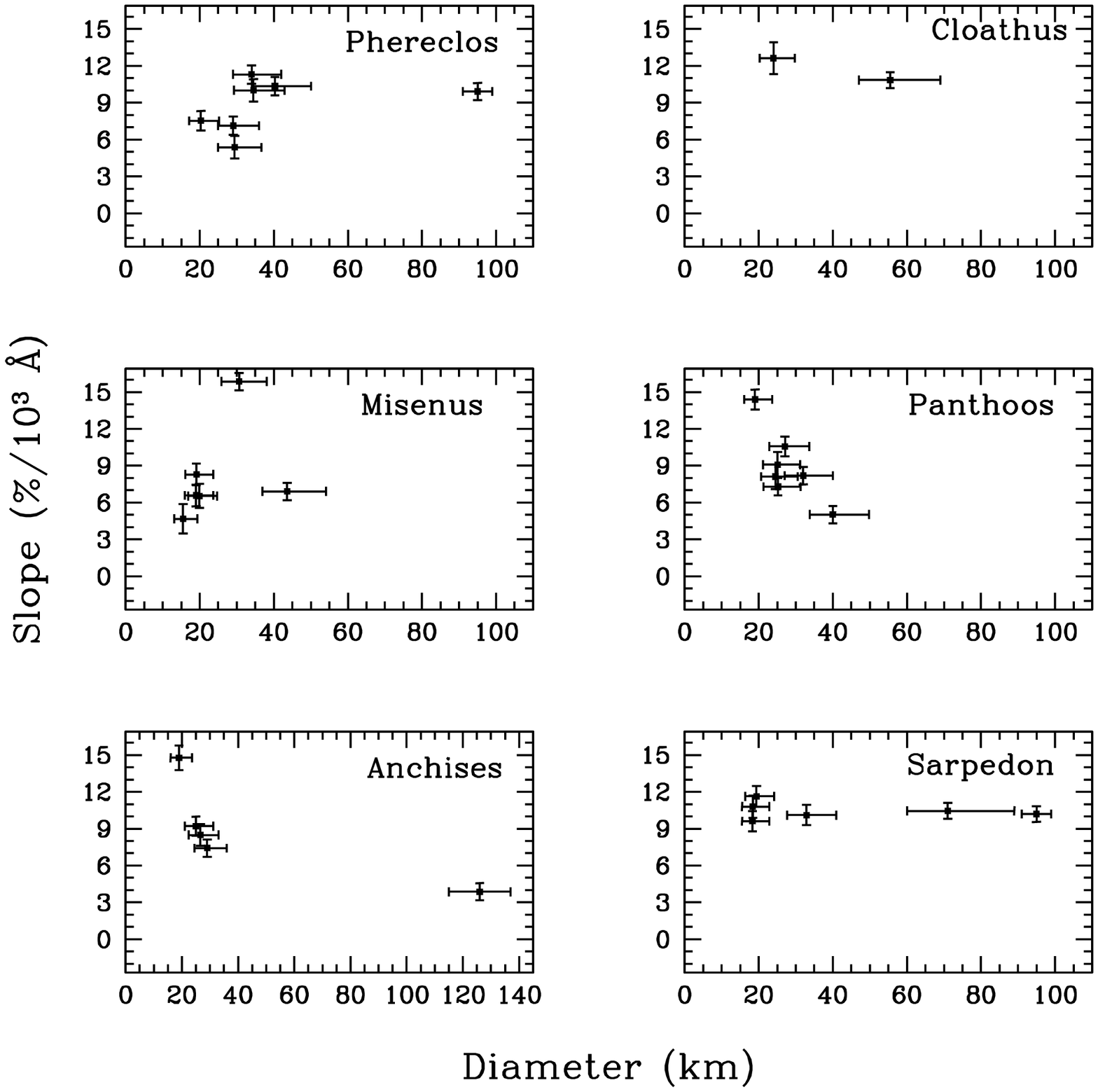,width=15truecm,angle=0}}
\caption{}
\label{fig7}
\end{figure}

\begin{figure}
\centerline{\psfig{file=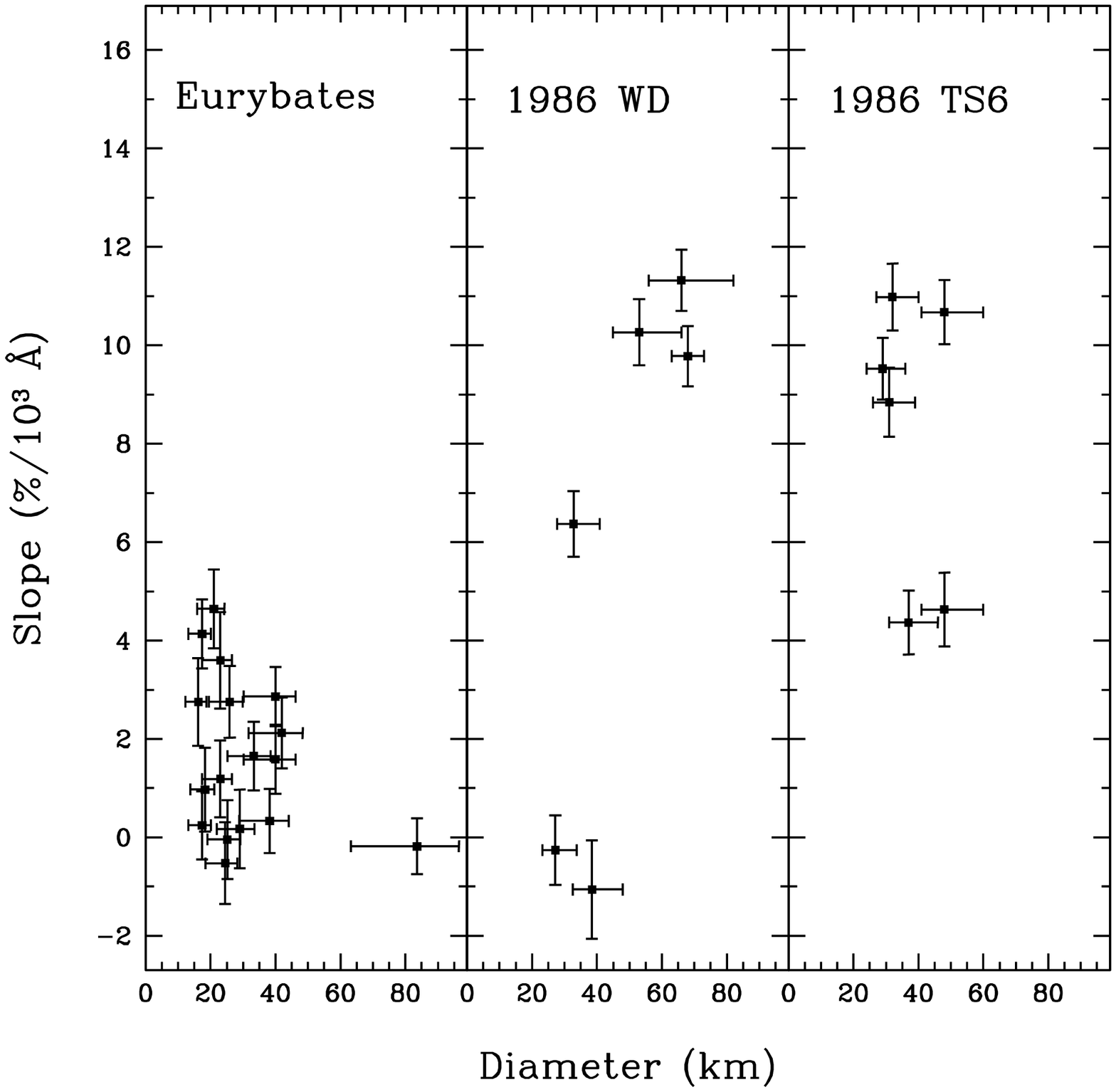,width=15truecm,angle=0}}
\caption{}
\label{fig8}
\end{figure}

We investigated 6 out of 17 members of the 4035 1986~WD family that is
dynamically defined at a cutoff of 130 m/s (Fig.~\ref{fig6} and
Table~\ref{spec_obsL4}). Three of our targets (4035, 6545 and 11351)
were already observed by Dotto et al. (2006): for 6545 and 11351 there
is a good consistency between our spectra and those already published.
4035 was observed also by Bendjoya et al. (2004): all the spectra are
featureless, but Bendjoya et al. (2004) obtain a slope of 8.8 $\% /
10^3$\AA, comparable to the one here presented, while Dotto et
al. (2006) found a higher value (see Table~\ref{spec_L4}).  This could
be interpreted as due to the different rotational phases seen in the
three observations, and could indicate some inhomogeneities on the
surface of 4035.

The observed family members show heterogeneous behaviors
(Fig.~\ref{fig8}), with spectral slopes ranging from neutral values
for the smaller members (24341 and 14707) to reddish ones for the 3
members with size bigger than 50 km (4035, 6545, and 11351).  For this
family, it seems that a size-slope relationship exists, with smaller
members having solar colors and spectral slopes increasing with the
object' sizes.

\subsubsection{1986~TS6}\label{sect:1986ts6}

The 1986~TS6 family includes 20 objects at a cut-off of 100 m/s. 
We present new spectroscopy and photometry of a single member, 12921 1998 WZ5 (Fig.~\ref{fig6}). The spectrum we present here is flat and featureless, with a spectral slope of $4.6\pm0.8$\%/$10^3$\AA.
Dotto et al. (2006) presented a spectrum obtained a month after our data (in May 2003) that has a very similar spectral slope $3.7\pm0.8$\%/$10^3$\AA. Previously, 12917 1998 TG16, 13463 Antiphos, 12921 1998 WZ5, 15535 2000 AT177, 20738 1999 XG191, and 24390 2000 AD177 were included in the Makhoan family. Refined proper elements now place all of these bodies in the 1986 TS6 family. \\

In Fig.~\ref{fig8} we report the spectral slopes vs. estimated
diameters of the 6 observed members. The family shows different
spectral slopes with the presence of both P--type (12921 and 13463) and
D--type asteroids (12917, 15535, 20738, and 24390).  Due to the very
similar diameters, a slope-size relationship is not found.

[FIGURE 7 AND 8]

%==============================================================================
%
\section{Discussion} 
%
%==============================================================================

The spectra of Jupiter Trojan members of dynamical families
show a range of spectral variation from C-- to D--type asteroids.  With
the exception of the L4 Eurybates family, all the observed objects have featureless spectra, and we cannot find any spectral bands which could help in the identification of minerals present on their surfaces.  The lack of detection of any mineralogy diagnostic feature
might indicate the formation of a thick mantle on the Trojan surfaces.  Such a mantle could be formed by a
phase of cometary activity and/or by space weathering processes as
demonstrated by laboratory experiments on originally icy surfaces
(Moore et al., 1983; Thompson et al., 1987; Strazzulla et al., 1998;
Hudson \& Moore, 1999).  \\
A peculiar case is constituted by the Eurybates family, which shows a preponderance of C--type objects and a total absence of D--types.
Moreover, this is the only family in which some members exhibit
spectral features at wavelengths shorter than 5000--5200 \AA, most likely due to the intervalence charge transitions in materials containing oxidized iron (Vilas 1994).  

\subsection{Size {\it vs} spectral slope distribution:\\Individual families}

The plots of spectral slopes vs. diameters are shown in Fig.~\ref{fig7}
and \ref{fig8}. A relationship between spectral slopes and 
diameters seems to exist for only three of the nine families we studied. 
In the Anchises and Panthoos families, smaller objects have redder spectra, while for the 1986 WD
family larger objects have the redder spectra.

Moroz et al. (2004) have shown that ion irradiation on natural
complex hydrocarbons gradually neutralizes the spectral slopes of
these red organic solids. If the process studied by Moroz et al. (2004) occurred on the
surface of Jupiter Trojans, the objects having redder spectra have to
be younger than those characterized by bluish-neutral spectra. In this
scenario the largest and spectrally reddest objects of the 1986 WD
family could come from the interior of the parent body and expose fresh material. 
In the case of the Anchises and
Panthoos families the spectrally reddest members, being the smallest,
could come from the interior of the parent body, or alternatively
could be produced by more recent secondary fragmentations. 
In particular, small family members may be more
easily resurfaced, as significant collisions (an impactor
having a size greater than a few percent of the target), as well as seismic shaking and recoating by
fresh dust, may occur frequently at small sizes.

[FIGURE 9]

\subsection{Size {\it vs} slope distribution: \\
           The Trojan population as a whole}
\label{sect:size}

[TABLE 6]

\begin{figure}
\centerline{\psfig{file=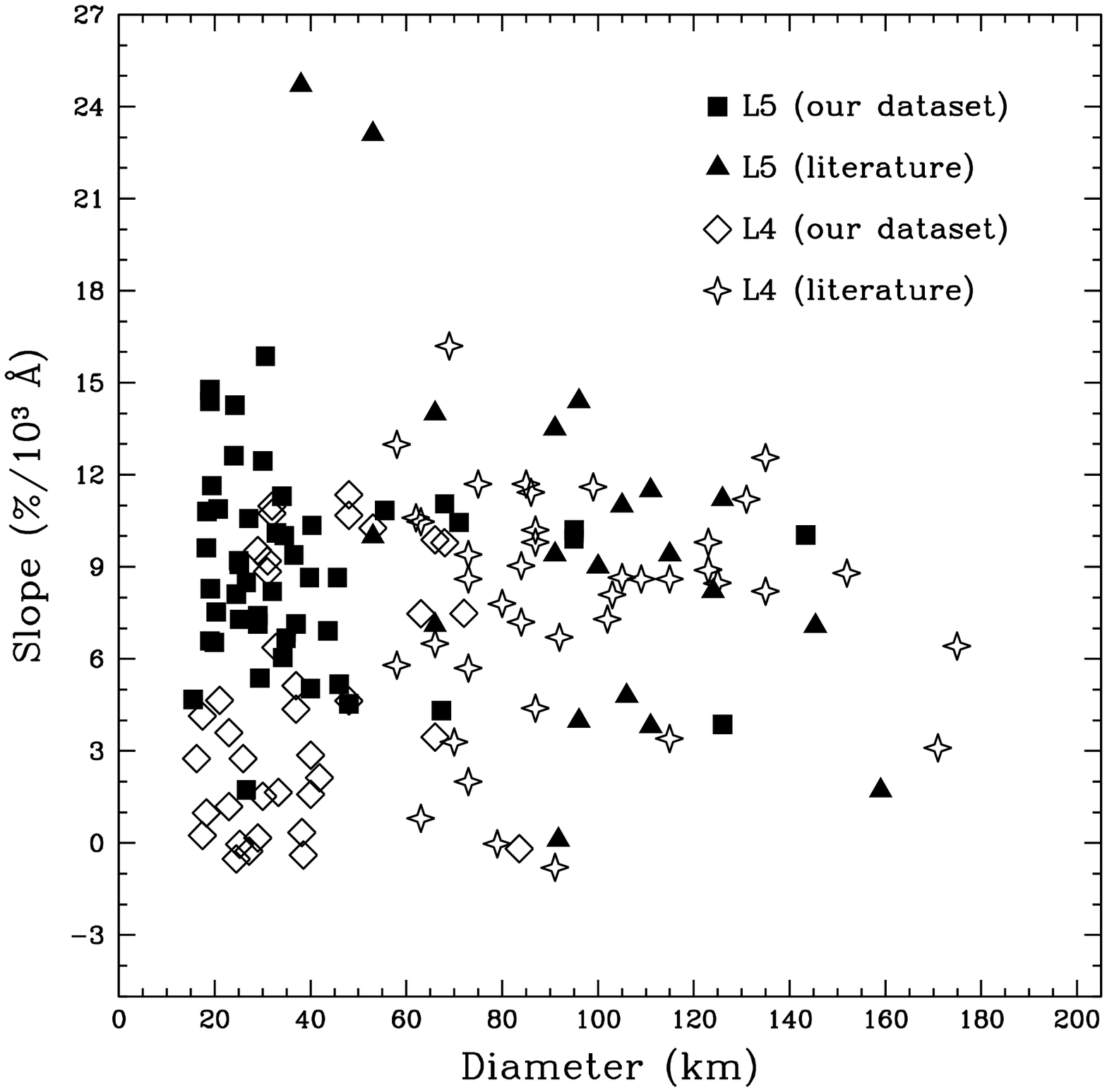,width=15truecm,angle=0}}
\caption{}
\label{fig9}
\end{figure}

As compared to the data available in literature, our sample strongly
contributed to the analysis of fainter and smaller Trojans, with
estimated diameters smaller than 50 km. Jewitt \& Luu (1990), analyzing a sample of 32 Trojans, found that the smaller objects were redder than the bigger ones.
However, our data play against the existence of a possible color-dimension trend.  
In fact, the spectral slope's range of the objects smaller than 50 km is similar to that of the 
larger Trojans, as shown in Fig.~\ref{fig9}.

The Eurybates family strongly contributes to the 
population of small spectrally {\it neutral} objects, filling the region
of bodies with mean diameter D$<$40 km and with spectral slopes smaller than 3 \%/$10^{3}$\AA. 

In order to carry out a complete analysis of the spectroscopic and
photometric characteristics of the whole available data set on Jupiter
Trojans, we considered all the visible spectra published in the
literature: Jewitt \& Luu (1990, 32 objects), Fitzimmons et al. (1994,
3 objects), Bendjoya et al. (2004, 34 objects), Fornasier et
al. (2004a, 26 L5 objects), and Dotto et al. (2006, 24 L4 Trojans).  We also add 
several Trojans spectra (11 L4 and 3 L5 Trojans) from the files
available on line (Planetary data System archive,
pdssbn.astro.umd.edu, and
www.daf.on.br/$\sim$lazzaro/S3OS2-Pub/s3os2.htm) from the SMASS I, SMASS
II and S3OS2 surveys (Xu et al., 1995; Bus \& Binzel, 2003; Lazzaro et
al., 2004). Including all these data, we compile a sample of 142
different Trojans, 68 belonging to the L5 cloud and 74 belonging to
the L4.  We performed the taxonomic classification of this enlarged
sample, on the basis of the Dahlgren and Lagerkvist (1995) scheme, by
analyzing spectral slopes computed in the range 5500-8000 \AA.
Different authors, of course, considered different spectral ranges for their own slope gradient evaluations: Jewitt \& Luu (1990) and Fitzimmons et al.
(1994) use the 4000-7400 \AA\ and Bendjoya et al. (2004, Table 2) used a
slightly different ranges around 5200-7500 \AA.  Since all the cited
papers show spectra with linear featureless trends, the different
wavelength ranges used for the spectral gradient computation by
Bendjoya et al. (2004) and Jewitt \& Luu (1990) are not expected to influence the obtained slopes.

In order to search for a dependency of the spectral
slope distribution with the size of the objects, all observations
(from this paper as well as from the literature) were combined. The
objects were isolated in 5 size bins (smaller than 25 km, 25--50 km,
50--75 km, 75-100 km and larger than 100 km). Each bin contains between
20 and 50 objects. These subsamples are large enough to be compared
using classical statistical tests: the t-test, which estimates if the
mean values are compatible, the f-test, which checks if the widths of
the distributions are compatible (even if they have different means),
and the KS test, which compares directly the full distributions. A
probability is computed for each test; a small probability indicates
that the tested distributions are {\em not} compatible, i.e. the
objects are not randomly extracted from the same population, while a
large probability value has no meaning (i.e. it is not possible to
assure that both samples come from the same population, we can
just say in that case that they are not incompatible).
In order to quantify the probability levels that we consider as
significant, the same tests were run on randomized distributions (see
Hainaut \& Delsanti 2002 for the method). Since probability
lower than 0.04-0.05 does not appear in these randomized
distributions, we consider that values smaller than 0.05 indicate a
significant incompatibility. \\
 Each sub-sample
was compared with the four others -- the results are summarized in
Table~\ref{tab:statSlopes}. The average slope of the 5 bins are all
compatible among each other. The only marginally significant result is
that the width of the slope distribution among the larger objects
(diam. $>100$ km) is narrower than that of all the smaller objects. \\
This narrower color distribution could be due to the aging processes affecting the surface of bigger objects, which are supposed to be older.
The wider color distribution of small members is possibly related to the different ages of their surfaces: some of them could be quite old, while some other could have been recently refreshed.

\subsection{Spectral slopes and L4/L5 Clouds}

[HERE FIGURES 10 AND 11]

Considering only the Trojan observations reported in this paper, the
average slope is 8.84$\pm$3.03\%/$10^{3}$\AA\ for the L5 population,
and 4.57$\pm$4.01\%/$10^{3}$\AA\ for the L4.

Considering now all the spectra available in the literature, the 68~L5
Trojans have an average slope of 9.15$\pm$4.19\%/$10^{3}$\AA, and the
78~L4 objects, 6.10$\pm$4.48\%/$10^3$\AA.  Performing the same
statistical tests as above, it appears that these two populations are
significantly different. In particular, the average slopes are
incompatible at the $10^{-5}$ level.\\
Nevertheless, as described in Section~\ref{sect:eurybates}, the
Eurybates family members have quite different spectral characteristics 
than the other objects and constitute a large subset of the whole sample. Indeed,
comparing their distribution with the whole populations, they are
found significantly different at the $10^{-10}$ level. In other words,
the Eurybates family members do not constitute a random subset of the other Trojans.

Once excluded the Eurybates family, the remaining 61 Trojans from the L4
swarm have an average slope of 7.33$\pm$4.24\%/$10^3$\AA. 
The very slight difference of average slope between the L5 and
remaining L4 objects is very marginally significant (probability of 1.6\%), and the
shape and width of the slope distributions are compatible with each
other.\\

\begin{figure}
\centerline{\psfig{file=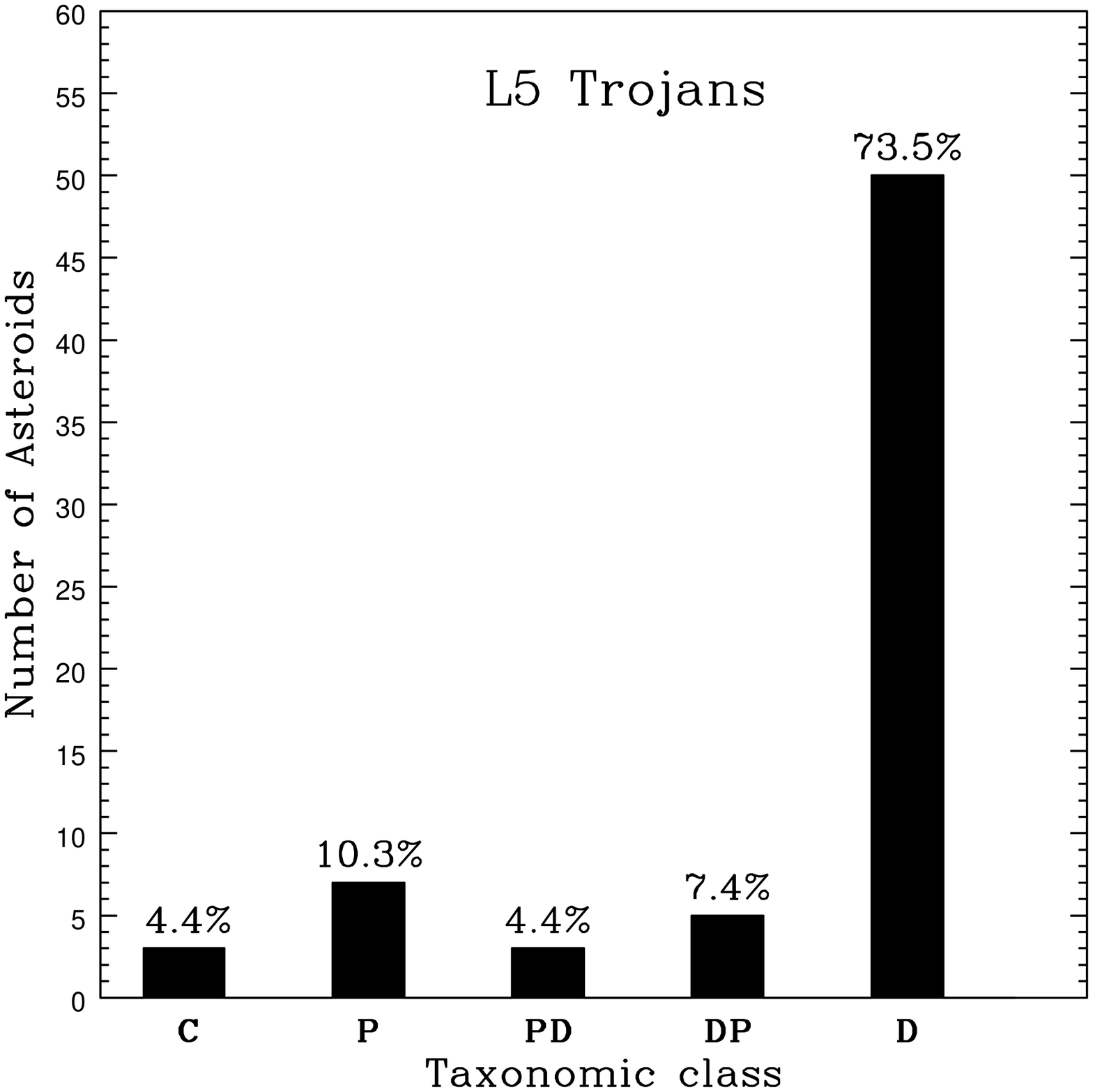,width=15truecm,angle=0}}
\caption{}
\label{fig10}
\end{figure}

The taxonomic classification we have performed shows that the majority
(73.5\%) of the observed L5 Trojans (Fig.~\ref{fig10}) are D--type (slope $>$ 7 \%/$10^{3}$ \AA) with featureless reddish
spectra, 11.8\% are DP/PD --type
(slope between 5 and 7 \%/$10^{3}$ \AA), 10.3\% are P--type, and
only 3 objects are classified as C--type (4.4\%).

\begin{figure}
\centerline{\psfig{file=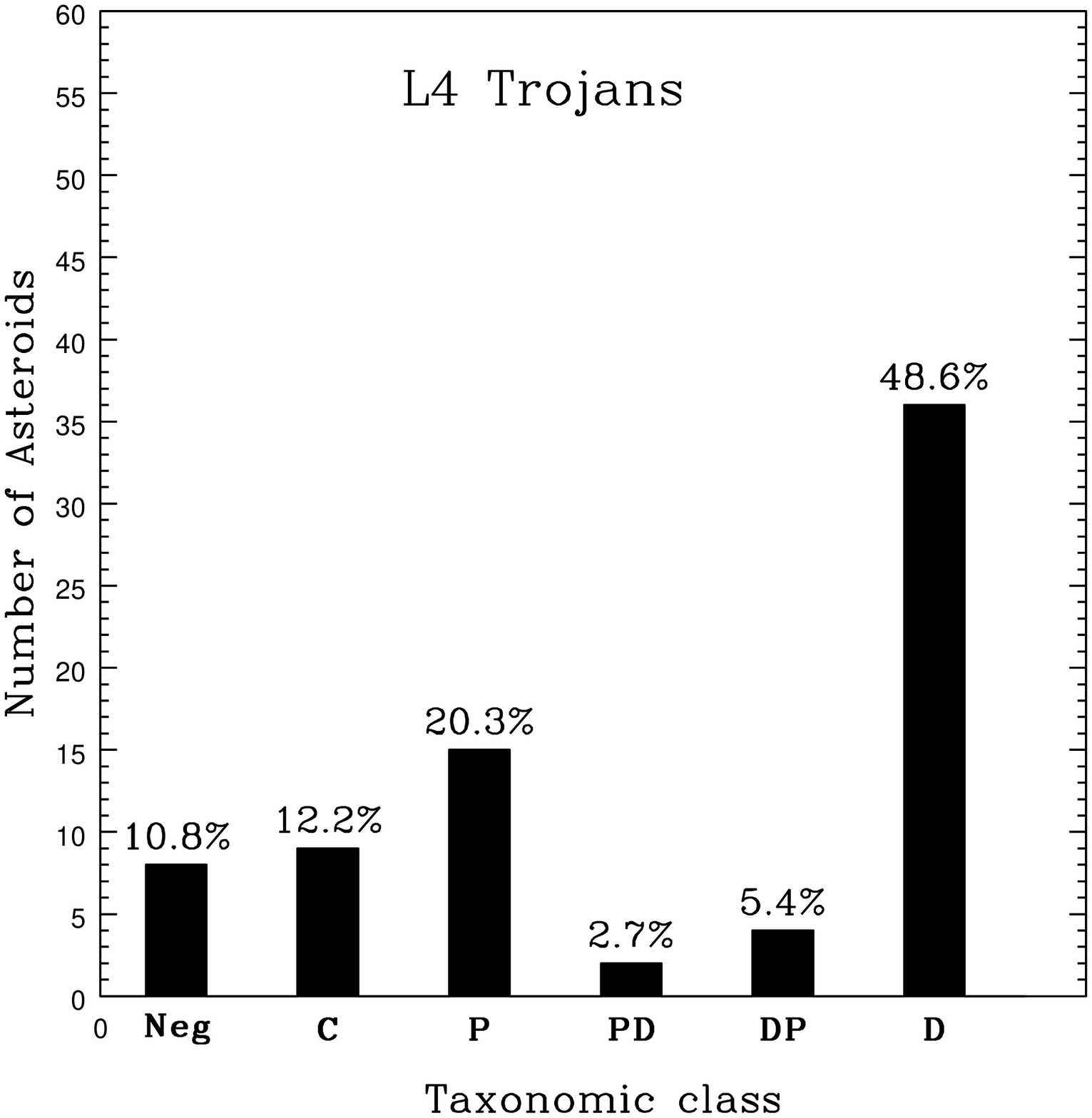,width=15truecm,angle=0}}
\caption{}
\label{fig11}
\end{figure}

In the L4 swarm (Fig.~\ref{fig11}), even though the D--type still
dominate the population (48.6\%), the spectral types are more
heterogeneous as compared to the L5 cloud, with a higher percentage
of neutral-bluish objects: 20.3\% are P--type, 8.1\% are DP/PD-type, 
12.2\% are C--type, and 10.8\% of the bodies have negative spectral slope. 
The higher percentage of C-- and P--type
as compared to the L5 swarm is strongly associated with the presence of
the very peculiar Eurybates family. Among 17 observed members 10 are
classified as C--types (among which 3 have negative spectral slopes)
and 7 are P--types. Considering the 57 asteroids that compose the
L4 cloud without the Eurybates family, we find percentages of P, and
PD/DP --types very similar to those of the L5 cloud (14.0\% and 10.5\%
respectively), a smaller percentage of D--types (63.2\%) and of the C--types (3.5\%), and the presence of a 8.8\% Trojans with negative spectral slopes.

The visible spectra of the Eurybates members are very similar to those of C--type main belt asteroids, Chiron-like Centaurs, and cometary nuclei. This similarity is compatible with three different scenarios: the family could have been produced by the fragmentation of a parent body very different from all the other Jupiter Trojans (in which case the origin of such a peculiar parent must still be assessed); this could be a very old family where space weathering processes have covered any differences in composition among the family members and flattened all the spectra; this could be a young family where space weathering processes occurred within time scales smaller than the age of the family.
In the last two cases the Eurybates family would give the first 
observational evidence of spectra flattened owing to space weathering 
processes. This would then imply, according to the results of Moroz et al. (2004), that its primordial composition was rich in complex hydrocarbons. \\
The knowledge of the age of the Eurybates family is therefore a 
fundamental step to investigate the nature and the origin of the parent 
body, and to assess the effect of space weathering processes on the 
surfaces of its members.

The present sample of Jupiter Trojans suggests a more heterogeneous
composition of the L4 swarm as compared to the L5 one.  As previously 
noted by Bendjoya et al. (2004), the L4 swarm contains a
higher percentage of C-- and P--type objects. 
This result is enhanced by members of the Eurybates family, but remains even 
when these family members are excluded. Moreover, the dynamical families
belonging to the L4 cloud are more robust than those of the L5 one,
surviving with densely populated clustering even at low relative
velocity cut-off. We therefore could argue that the L4 cloud is more collisionally active than the L5 swarm. Nevertheless, we still cannot intepret this in terms of the composition of the two populations, since we cannot exclude that as yet unobserved C-- and P--type families are present in the L5 cloud.

\subsection{Orbital Elements}

[HERE FIGURE 12 and TABLES 7 and 8 ]

\begin{figure}
 	 \epsfxsize16.cm\epsfbox{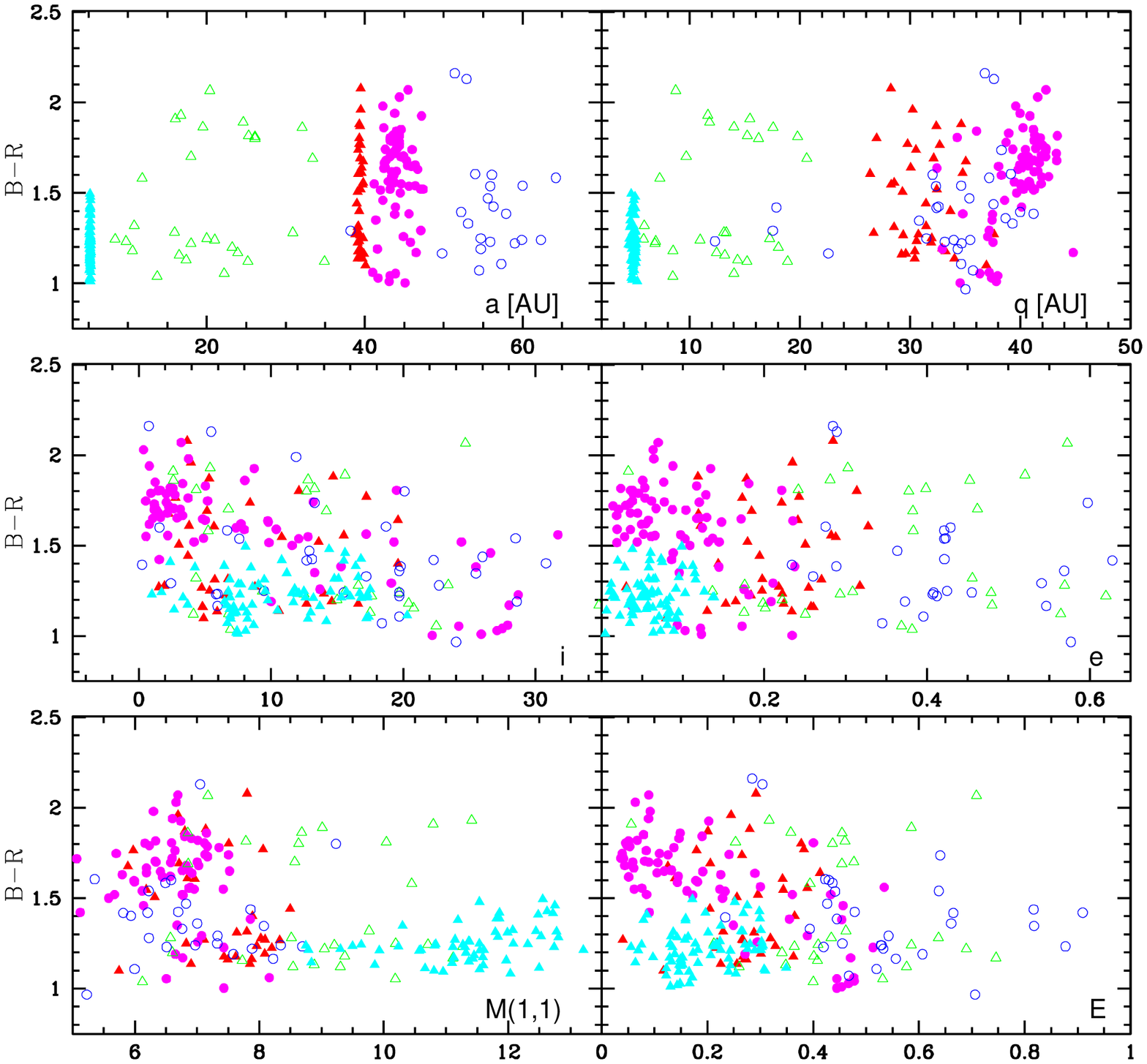}
 	 \caption{}
	 \label{fig:orbit} 
\end{figure}

We analyzed the spectral slope as a function of the Trojans' orbital
elements. As an illustration, Fig.~\ref{fig:orbit} shows the $B-R$
color distribution as a function of the orbital elements. In order to
investigate variations with orbital parameters, the Trojan population is
divided in 2 sub samples: those with the considered orbital element
lower than the median value, and those with the orbital element higher
than the median (by construction, the two subsamples have the same
size). Taking $a$ as an example, half the Trojans have $a<5.21$AU, and
half have $a$ larger than this value.

The mean color, the color dispersion, and the color distribution of
the 2 subsamples are compared using the three statistical tests mentioned
in Section~\ref{sect:size}. The method is discussed in details in
Hainaut \& Delsanti (2002).  The tests are repeated for all color and
spectral slope distributions.  The results are the following.

\begin{itemize}
\item $q$, perihelion distance: the color distribution of the Trojans
  with small $q$ is marginally broader than that of Trojans with larger
  $q$. This result is not very strong (5\%), and is dominated by the
  red-end of the visible wavelength. Removing the Eurybates from the
  sample maintains the result, at the same weak level.
%%%
\item $e$, eccentricity: the distribution shows a similar result, also at
  the weak 5\% significance. The objects with larger $e$ have broader
  color distribution than those with lower $e$. This result is
  entirely dominated by the Eurybates' contribution. 
%%%
\item $i$, inclination: objects with smaller inclination are significantly bluer
  than those with larger $i$. This result is observed
  at all wavelengths. It is worth noting that this is contrary to what is
  usually observed on other Minor Bodies in the Outer Solar System survey (MBOSSes), where objects with high $i$, or
  more generally, high excitation $E=\sqrt{e^2 + \sin^2i}$, are bluer (Hainaut \& Delsanti, 2002; Doressoundiram et al., 2005). This can also be visually appreciated in Fig.~\ref{fig:orbit}. This
  result is also completely dominated by the Eurybates'
  contribution. The non-Eurybates Trojans do not display this trend.
%%%
\item $E=\sqrt{e^2 + \sin^2i}$, orbital excitation: the objects with
  small $E$ are also significantly bluer than those with high $E$.
  This result is also completely dominated by the Eurybates'
  contribution. The non-Eurybates Trojans do not display this trend.
%%%
\end{itemize}

In summary this analysis shows that the Eurybates sub-sample of the Trojans is well separated in orbital elements and in colors. 

For the other Minor Bodies in the outer Solar System, the relation
between color and inclination--orbital excitation (objects with a
higher orbital excitation tend to be bluer) is interpreted as a
relation between excitation and surface aging/rejuvenating processes (Doressoudiram et al., 2005).
The Eurybates family has low excitation and neutral-blue colors,
suggesting that the aging/rejuvenating processes
affecting them are different from the other objects. This could be due to different surface compositions, different irradiation processes, or different collisional properties -- which would be
natural for a collisional family.

%==============================================================================
%
   \section{Comparison with other outer Solar System minor bodied}
%
%==============================================================================

%%%oli
\subsection{Introduction and methods}

[HERE FIGURES 13 AND 14]

\begin{figure}
   \epsfxsize15cm\epsfbox{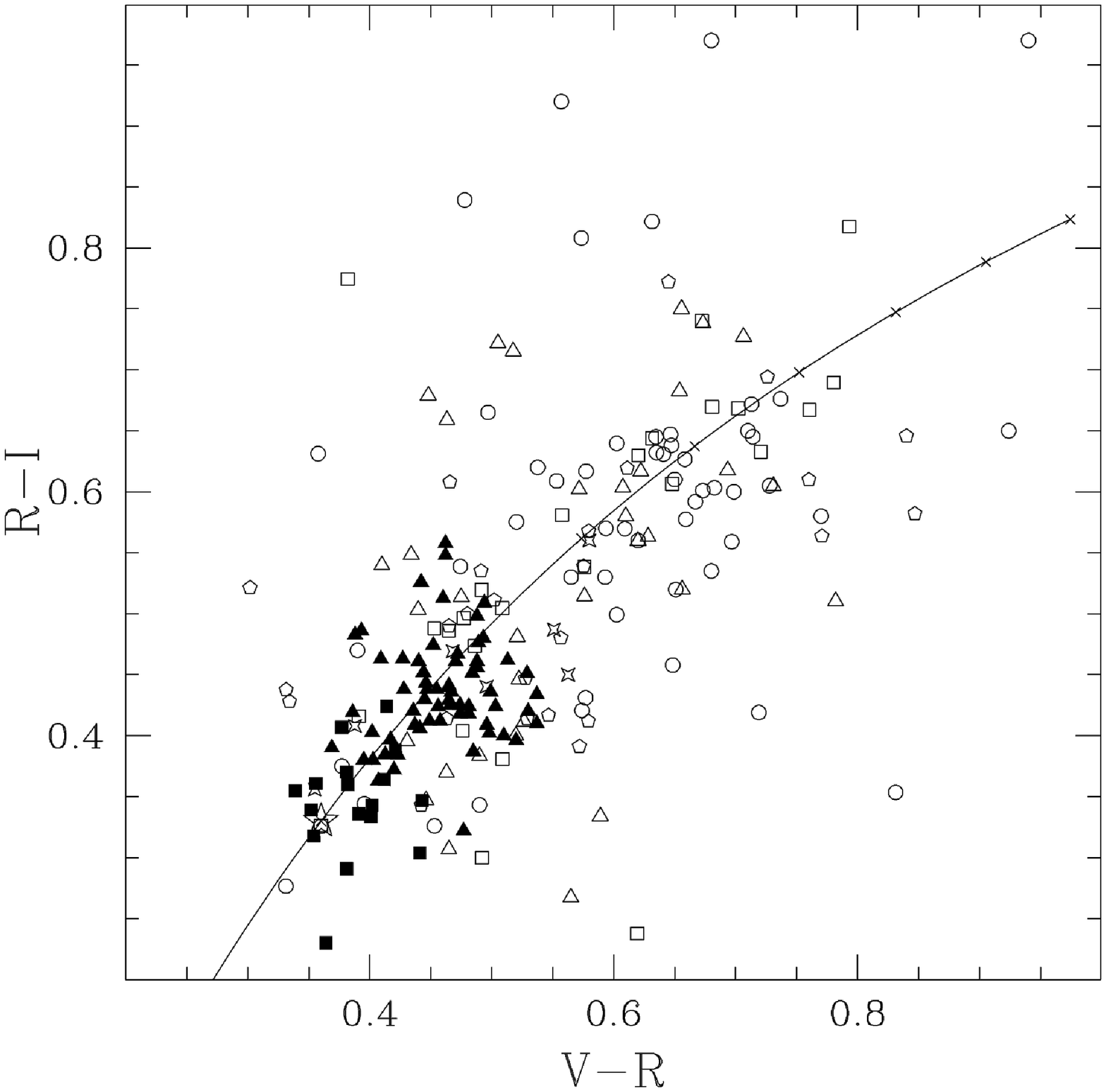}\\
   \caption{}
   \label{fig:vrri}
\end{figure}

\begin{figure}
   \epsfxsize6.7cm\epsfbox{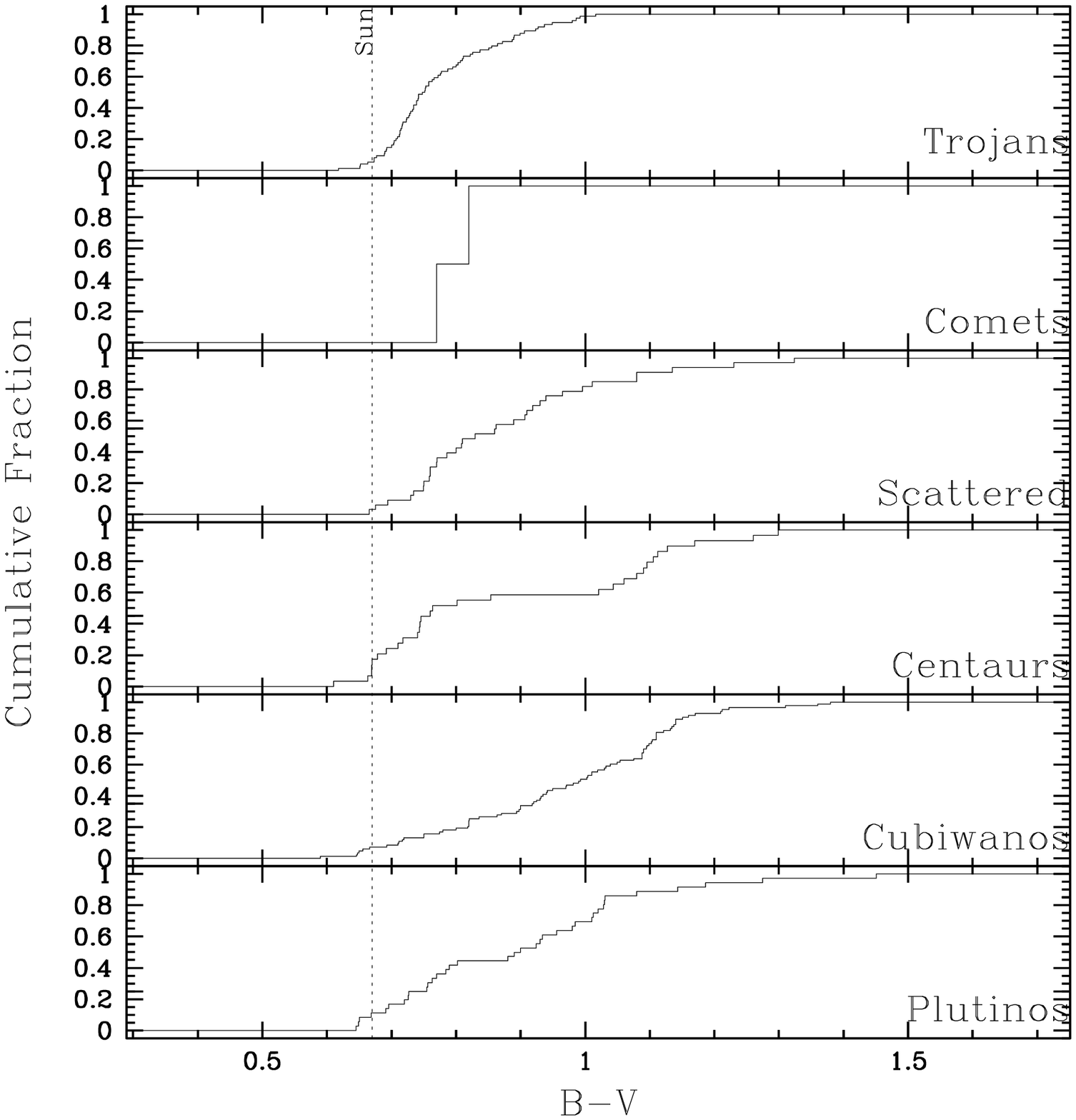}
   \epsfxsize7cm\epsfbox{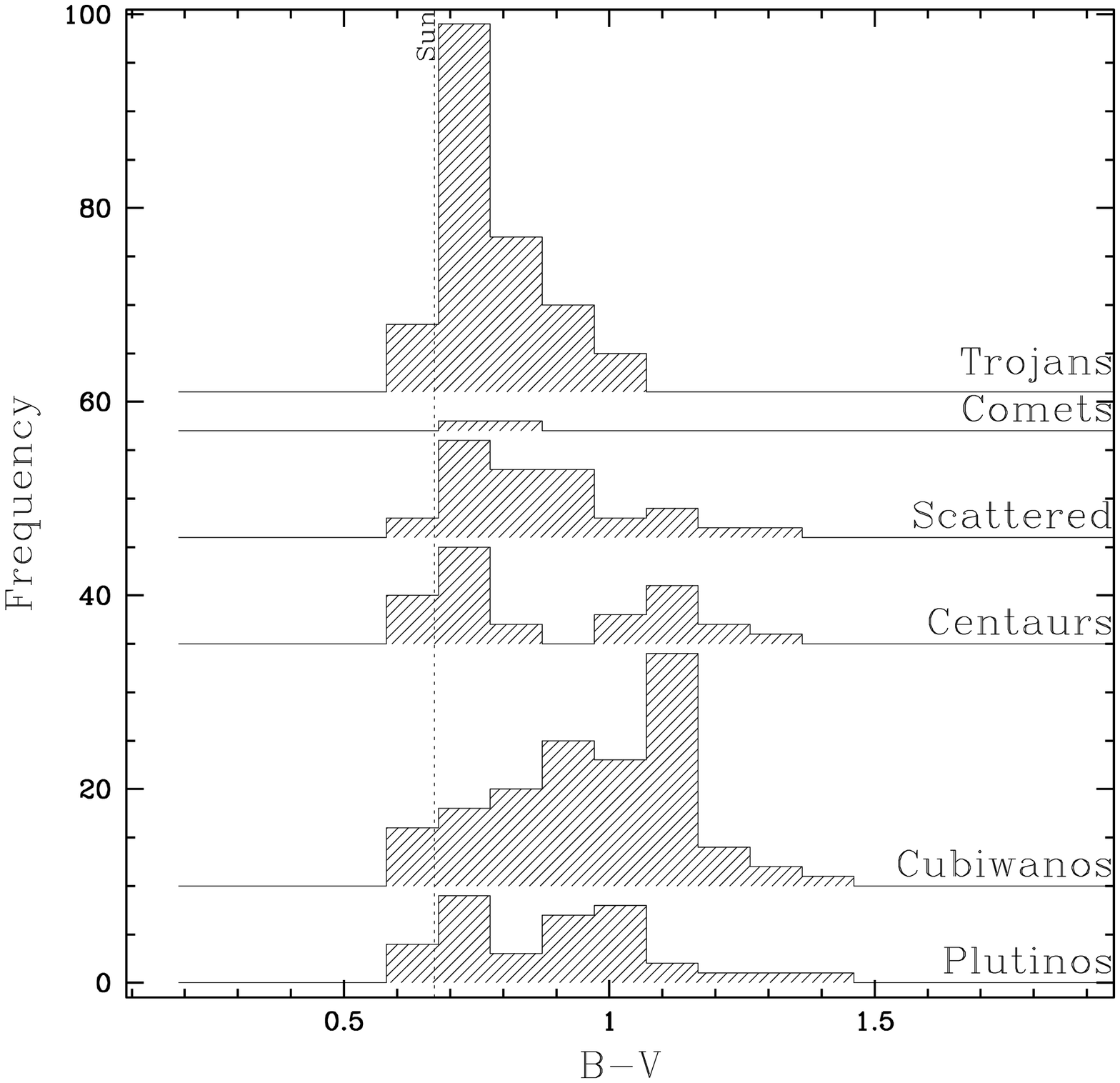}  \\

   \epsfxsize6.7cm\epsfbox{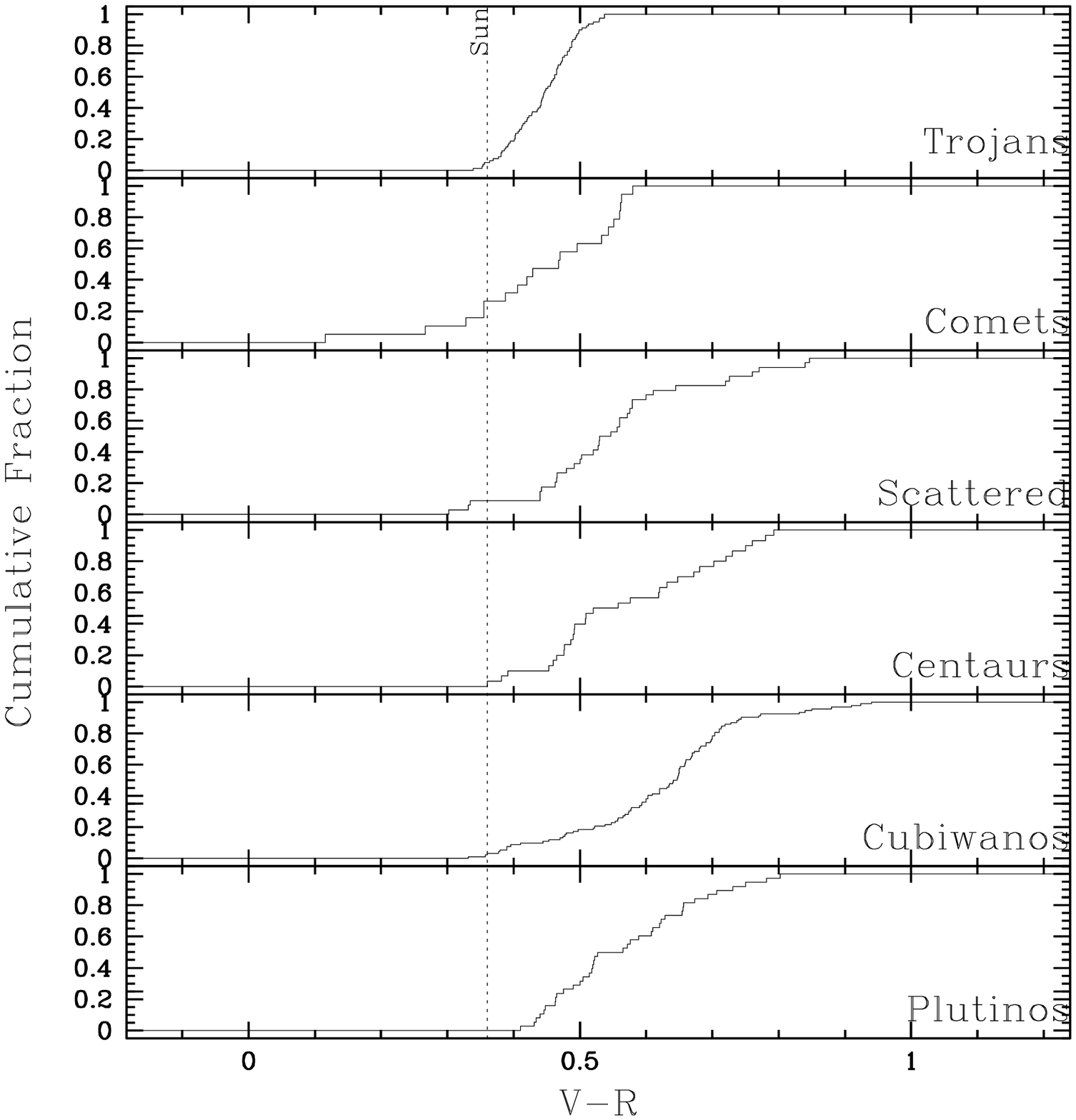}
   \epsfxsize7cm\epsfbox{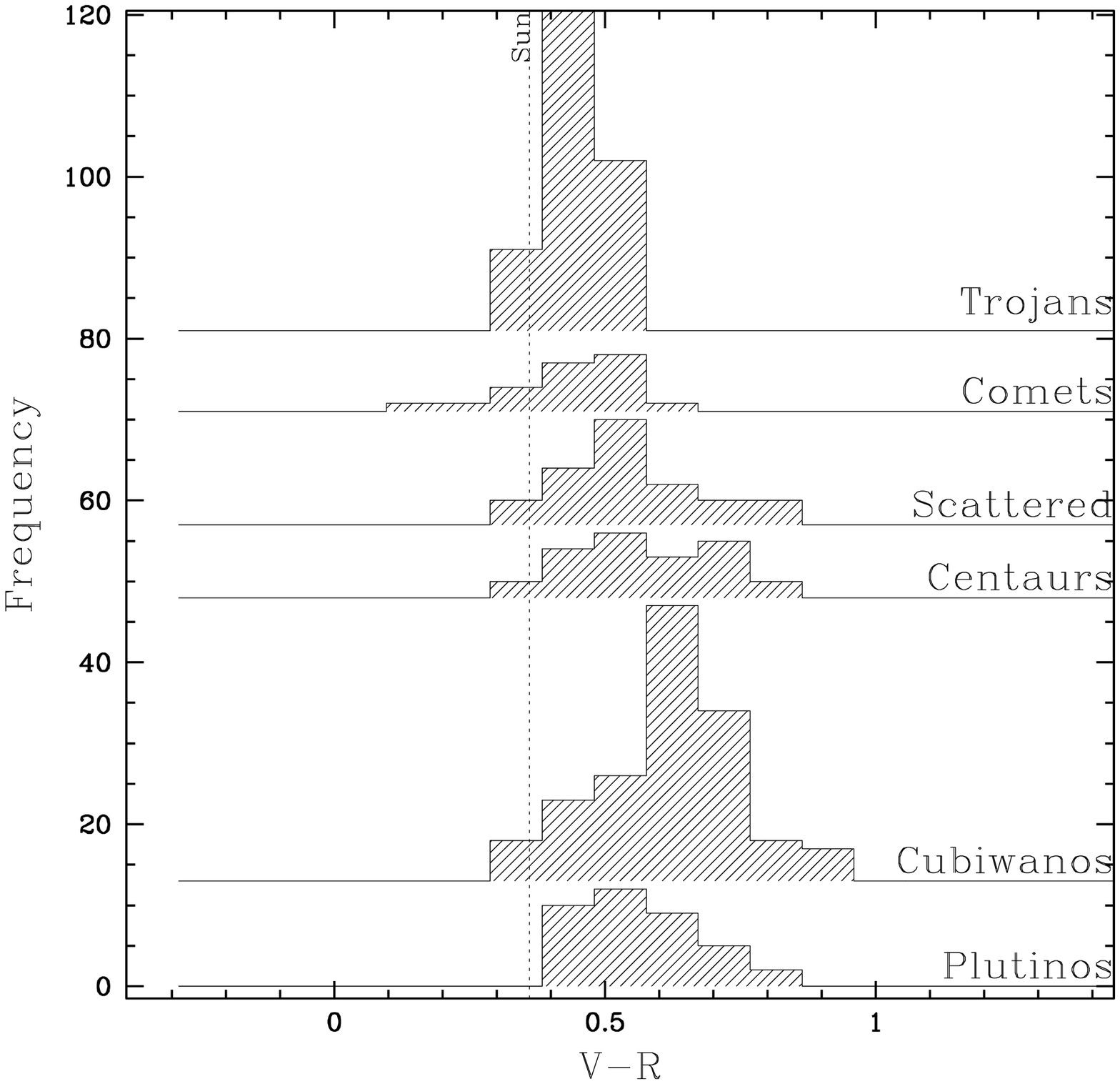}  \\

   \epsfxsize6.7cm\epsfbox{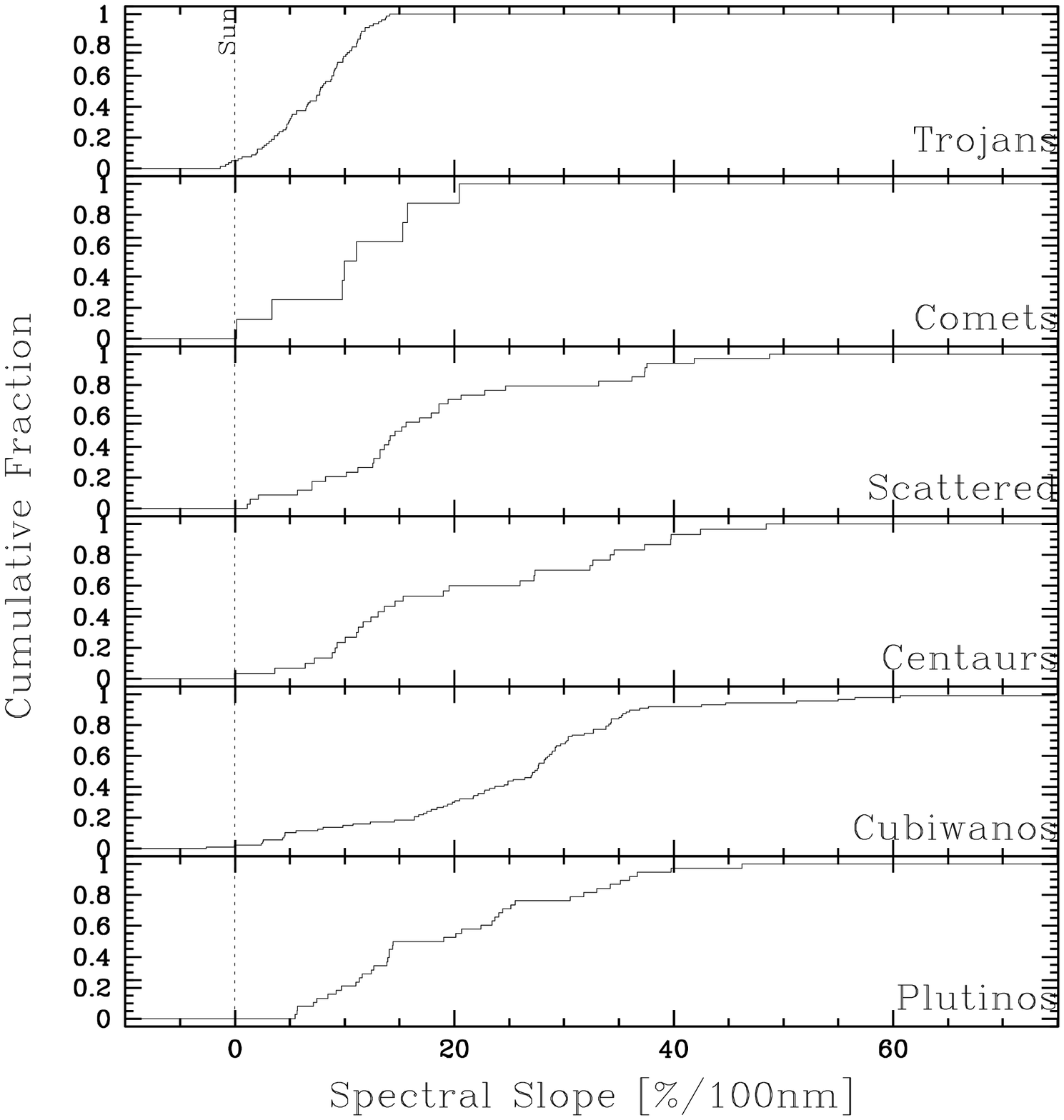}
   \epsfxsize7cm\epsfbox{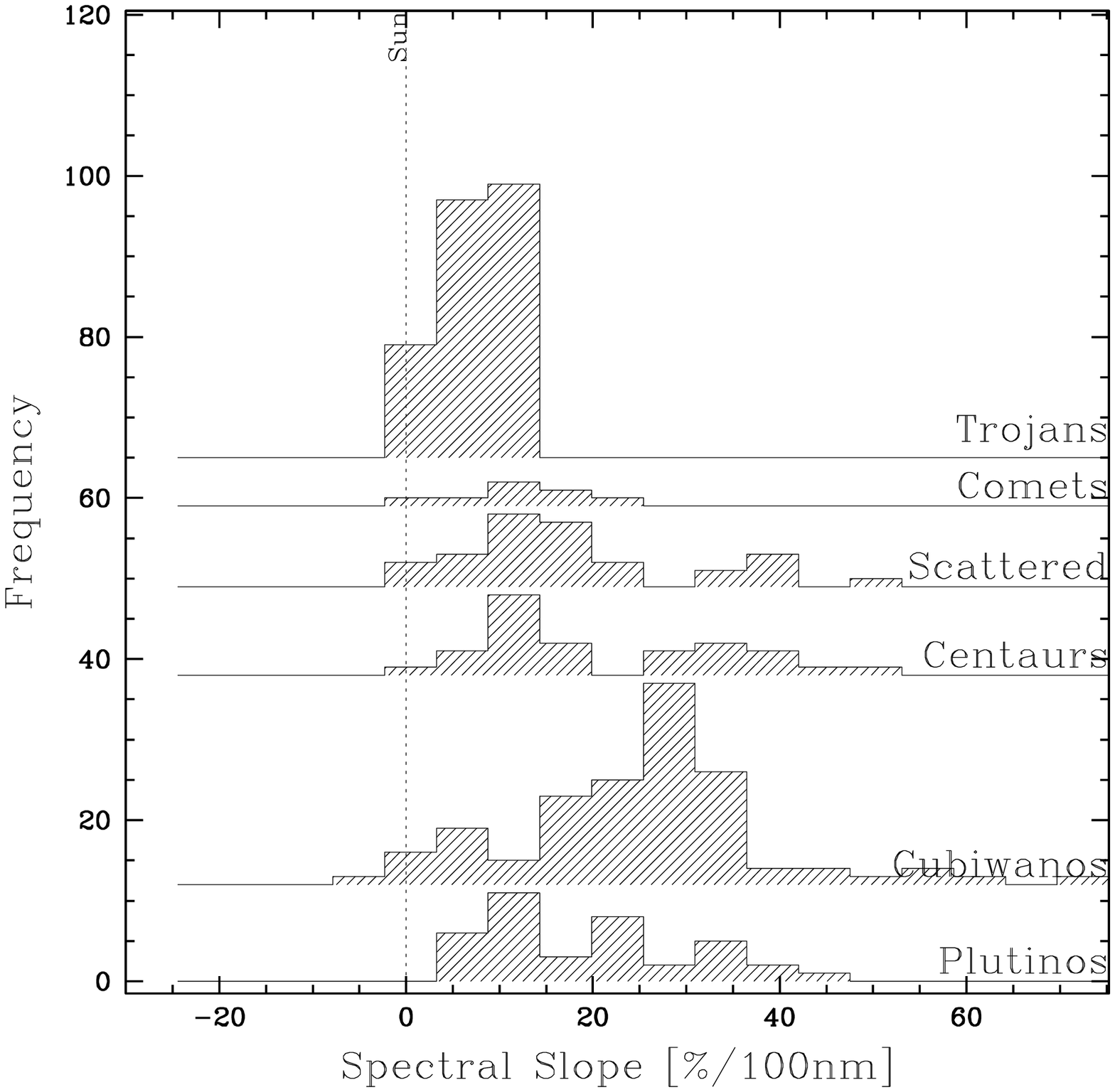}  \\

   \caption{}
   \label{fig:colorCPF}
 \end{figure}

The statistical tests set described in section 4.2 has also been
applied to compare the colors and the spectral slopes distribution of the
Trojans with those of the other minor bodies in the outer Solar System
taken from the updated, on-line version of the Hainaut \& Delsanti
(2002) database. Figure~\ref{fig:vrri}, as an example, displays the 
(R-I) vs (V-R) diagrams, while Fig.~\ref{fig:colorCPF}
shows the (B-V) and (V-R) color distributions, as well as the spectral slope distribution of the different classes of
objects.  The tests were performed on all the color indices derived from filters in the visible (UBVRI) and near infrared range (JHK) but in
Table~\ref{tab:average} and~\ref{tab:tests} we summarize the
most significant results.

In order to ``calibrate'' the significant probabilities, additional
artificial classes are also compared: first, the objects which have an
even internal number in the database with the odd ones. As this
internal number is purely arbitrary, both classes are statistically
indistinguishable. The other tested pair is the objects with a
``1999'' designation versus the others.  
Again, this selection criterion is arbitrary, so the pseudo-classes it
generates are sub-sample of the total population, and should be
indistinguishable. However, as many more objects have been discovered
in all the other years than during that specific year, the size of
these sub-samples are very different. This permits us to estimate the
sensitivity of the tests on sample of very different sizes.
Some of the tests found the arbitrary populations
incompatible at the $5\%$ level, so we use 0.5\% as a conservative threshold for statistical significance of
the distribution incompatibility

\subsection{Results}

Table~\ref{tab:average} and Fig.~\ref{fig:colorCPF} clearly show that
the Trojans' colors distribution is different as compare to that of Centaurs, TNOs and comets. Trojans are at the same time
bluer, and their distribution is narrower than all the other
populations. Using the statistical tests (see Table~\ref{tab:tests}),
we can confirm the significance of these results.

\begin{itemize}

\item The average colors of the Trojans are significantly different
  from those of all the other classes of objects (t-test), with the
  notable exception of the short period comet nuclei. Refining the
  test to the Eurybates/non-Eurybates, it appears that the Eurybates
  have marginally different mean colors, while the non-Eurybates
  average colors are indistinguishable from those of the comets.

\item Considering the full shape of the distribution (KS test), we 
obtain the same results: the Trojans colors distributions are
  significantly different from those of all the other classes, with
  the exception of the SP comets, which are compatible. Again, 
this result becomes stronger separating the Eurybates: their
  distributions are different from those of the comets, while the
  non-Eurybates ones are indistinguishable. 

\item The results when considering the widths of the color 
 distributions (f-test) are slightly different. Classes of objects with 
different mean colors could still
  have the same distribution width. This could suggest that a similar
  process (causing the width of the distribution) is in action, but
  reached a different equilibrium point (resulting in different mean
  values). This time, all classes are incompatible with the Trojans,
  including the comets, with strong statistical significance.
\end{itemize}

In order to further explore possible similarities between Trojans and
other classes, the comparisons were also performed with the neutral
Centaurs. These were selected with $S<20$\%/$10^{3}$\AA); this
cut-off line falls in the gap between the "neutral" and "red" Centaurs (Peixinho et al., 2003, Fornasier et al., 2004b).

The t-Test (mean color) only reveals a very moderate incompatibility
between the Trojans and neutral Centaurs, at the 5\% level, i.e. only
marginally significant. On the other hand, the f-Test gives some
strong incompatibilities in various colors (moderate in $B-V$ and $H-K$,
very strong in $R-I$), but the two populations are compatible for most 
of the other colors. Similarly, only the $R-I$ KS-test reveals a strong
incompatibility. It should also be noted that only 18 neutral Centaurs
are known in the database. In summary, while the Trojans and neutral Centaurs have fairly similar mean colors, their color distributions
are also different.

\section{Conclusions}

From 2002, we carried out a spectroscopic and photometric survey of
Jupiter Trojans, with the aim of investigating the members of
dynamical families. 
In this paper we present new data on 47 objects belonging to several dynamical families: Anchises (5 members), Cloanthus (2
members), Misenus (6 members), Phereclos (3 members), Sarpedon (2
members) and Panthoos (5 members) from the L5 swarm; Eurybates (17
members), 1986 WD (6 members), and Menelaus (1 member) for the L4 swarm. Together with the data
already published by Fornasier et al. (2004a) and Dotto et al. (2006),
taken within the same observing program, we have a total sample of 80
Trojans, the largest homogeneous data set available to date on these
primitive asteroids. The main results coming from the observations
presented here, and from the analysis including previously published visible 
spectra of Trojans, are the following:

\begin{itemize}

\item Trojans' visible spectra are mostly featureless. However, some
members of the Eurybates family show a UV drop-off in reflectivity for wavelength shorter than 5000--5200 \AA\ that is possibly due to intervalence charge transfer transitions (IVCT) in oxidized iron.

\item The L4 Eurybates family strongly differs from all the other families in that it is dominated by C-- and P--type asteroids. Also its spectral slope distribution is significantly different when compared to that of the other Trojans (at the $10^{-10}$ level). \\ 
This family is
very peculiar and is dynamically very strong, as it survives also at
a very stringent cutoff (40 m/s). Further observations in the
near-infrared region are strongly encouraged to look for possible
absorption features due to water ice or to material that experienced
aqueous alteration.

\item The average spectral slope for the L5 Trojans is
  9.15$\pm$4.19\%/$10^{3}$\AA, and 6.10$\pm$4.48\%/$10^3$\AA\  for the
  L4 objects. Excluding the Eurybates, the L4 average slope values
  becomes 7.33$\pm$4.24\%/$10^3$\AA. The slope distributions of the L5 and
  of the non-Eurybates L4 are indistinguishable.

\item Both L4 and L5 clouds are dominated by D--type asteroids, but
the L4 swarm has an higher presence of C-- and P--type asteroids, even when the Eurybates family is excluded, and
appears more heterogeneous in composition as compared to the L5 one.

\item We do not find any size versus spectral slope relationship
inside the whole Trojans population.

\item The Trojans with higher orbital inclination are significantly
redder than those with lower $i$. While this trend is the opposite of
that observed for other distant minor bodies, this effect is entirely
dominated by the Eurybates family.

\item Comparing the Trojans colors with those of other distant minor
  bodies, they are the bluest of all classes, and their colors
  distribution is the narrowest. This difference is mostly due to the Eurybates family. In fact, if we consider only the Trojan population without the Eurybates members, their average colors and overall distributions are not distinguishable from that of the short period comets. However, the widths of their color
  distributions are not compatible. The similarity in the overall color distributions might be caused by the small size of the short period comet sample rather
  than by a physical analogy. The Trojans average colors
  are also fairly similar to those of the neutral Centaurs, but the overall
  distributions are not compatible. 
\end{itemize}  

After this study, we have to conclude that Trojans
  have peculiar characteristics very different from those of all the other
  populations of the outer Solar System.\\
Unfortunately, we still cannot assess if this is due to differences in
the physical nature, or in the aging/rejuvenating 
processes which modified the surface materials in different way at different solar distances. 
Further observations, mainly in V+NIR spectroscopy and polarimetry, are 
absolutely needed to better investigate the nature of Jupiter Trojans and 
to definitively assess if a genetical link might exist with Trans-Neptunian 
Objects, Centaurs and short period comets.

{\bf Acknowledgments} \\
We thank Beaug\'e and Roig for kindly providing us with updated 
Trojan family list, and R.P. Binzel and J.P. Emery for their useful comments in the reviewing process.

\bigskip

{\bf References} \\

Barucci, M. A., Lazzarin, M., Owen, T., Barbieri, C., Fulchignoni, M., 1994. Near--infrared spectroscopy of dark asteroids. Icarus 110, 287-291. \\

Beaug\'e, C., Roig, F., 2001.  A Semianalytical Model for the Motion of the Trojan Asteroids: Proper Elements and Families. Icarus 53, 391-415. \\

Bendjoya, P., Cellino, A., Di Martino, M., Saba, L., 2004. Spectroscopic observations of Jupiter Trojans. Icarus 168, 374-384. \\

Binzel, R. P., Sauter, L. M., 1992. Trojan, Hilda, and Cybele asteroids - New lightcurve observations and analysis. Icarus 95, 222-238. \\

Bowell, E., Hapke, B., Domingue, D., Lumme, K., Peltoniemi, J., Harris, A.W.,
2003.
Application of photometric models to asteroids. In Asteroids II (R.P Binzel,
T. Gehrels, M.S. Matthews, eds)
Univ. of Arizona Press, Tucson, pp. 524--556. \\

Bus, S. J., Binzel, R.P., 2003. Phase II of the Small Main-Belt Asteroid Spectroscopic Survey. The Observations. Icarus 158, 106--145. \\

Dahlgren, M., Lagerkvist, C. I., 1995. A study of Hilda asteroids. I. CCD spectroscopy of Hilda asteroids. Astron. Astrophys. 302, 907-914. \\

Dell'Oro, A., Marzari, P., Paolicchi F., Dotto, E., Vanzani, V., 1998. Trojan collision probability: a statistical approach. Astron. Astrophys. 339, 272-277. \\

Doressoundiram, A., Peixinho, N., Doucet, C., Mousis, O., Barucci, M.~A., Petit, J.~M.,
Veillet, C., 2005. The Meudon Multicolor Survey (2MS) of Centaurs and
trans-neptunian objects: extended dataset and status on the correlations
reported. Icarus 174, 90--104. \\

Dotto, E., Fornasier, S., Barucci, M. A., Licandro, J., Boehnhardt, H., Hainaut, O., Marzari, F., de Bergh, C., De Luise, F., 2006. The surface composition
of Jupiter Trojans: Visible and Near--Infrared Survey of Dynamical Families. Icarus 183, 420-434 \\  

Dumas, C., Owen, T., Barucci, M. A., 1998. Near-Infrared Spectroscopy of Low-Albedo 
Surfaces of the Solar System: Search for the Spectral Signature of Dark Material. 
Icarus 133, 221-232. \\
            
Emery, J. P., Brown, R. H., 2003. Constraints on the surface composition of Trojan asteroids from near-infrared (0.8-4.0 $\mu$m) spectroscopy. Icarus 164, 104-121.\\

Emery, J. P., Brown, R. H., 2004. The surface composition of Trojan asteroids: constraints set by scattering theory. Icarus 170, 131-152.\\

Emery, J. P., Cruikshank, D. P., Van Cleve, J., 2006. Thermal emission spectroscopy (5.2 38 $\mu$m of three Trojan asteroids with the Spitzer Space Telescope: Detection of fine-grained silicates. Icarus 182, 496-512. \\

Fernandez Y. R., Sheppard, S. S., Jewitt, D. C., 2003. The Albedo Distribution of Jovian Trojan Asteroids. Astron. J. 126, 1563-1574. \\

Fitzsimmons, A., Dahlgren, M., Lagerkvist, C. I., Magnusson, P., Williams, I. P., 1994. A spectroscopic survey of D-type asteroids. Astron. Astrophys.  282, 634-642. \\

Fornasier, S., Lazzarin, M., Barbieri, C., Barucci, M. A., 1999. Spectroscopic comparison of aqueous altered asteroids with CM2 carbonaceous chondrite meteorites.  Astron. Astrophys.  135, 65-73  \\ 

Fornasier, S., Dotto, E., Marzari, F., Barucci, M.A., Boehnhardt, H., Hainaut, O.,
de Bergh, C., 2004a. Visible spectroscopic and photometric survey of L5 Trojans
: investigation of dynamical families. Icarus, 172,  221--232. \\

Fornasier, S., Doressoundiram, A., Tozzi, G. P., Barucci, M. A., Boehnhardt, H., 
de Bergh, C., Delsanti A., Davies, J., Dotto, E., 2004b. ESO Large Program on Physical 
Studies of Trans-Neptunian Objects and Centaurs: final results of the visible 
spectroscopic observations. Astron. Astrophys. 421, 353-363. \\

Hainaut, O. R., Delsanti, A. C., 2002. Colors of Minor Bodies in the Outer Solar System. A statistical analysis.  Astron. Astroph. 389, 641-664.   \\

Hudson, R.L., Moore, M.H. 1999. Laboratory Studies of the Formation of
Methanol and Other Organic Molecules by Water+Carbon Monoxide Radiolysis:
Relevance to Comets, Icy Satellites, and Interstellar Ices.
Icarus 140, 451-461.\\

Jewitt, D. C., Luu, J. X., 1990. CCD spectra of asteroids. II - The Trojans as spectral analogs of cometary nuclei. Astron. J. 100, 933-944. \\

Jewitt, D. C., Trujillo, C. A., Luu, J. X., 2000. Population and Size Distribution of Small Jovian Trojan Asteroids. Astron. J. 120, 1140-1147 \\

Landolt, A. U., 1992. UBVRI photometric standard stars in the magnitude range 11.5--16.0 around the celestial equator. Astron. J
. 104, 340-371, 436-491. \\

Lazzaro, D., Angeli, C. A., Carvano, J. M., Moth\'e-Diniz, T., Duffard, R., Florczak, M., 2004. S$^{3}$OS$^{2}$: the visible spectroscopic survey of 820 asteroids. Icarus 172, 179--220.\\

Levison, H., Shoemaker, E. M., Shoemaker, C. S., 1997. The dispersal of the Trojan asteroid swarm. Nature 385, 42-44. \\

Marzari, F., Farinella, P., Davis, D. R., Scholl, H., Campo Bagatin, A., 1997. 
Collisional Evolution of Trojan Asteroids. Icarus 125, 39-49. \\

Marzari, F., Scholl, H., 1998a. Capture of Trojans by a Growing Proto-Jupiter. Icarus 131, 41-51.\\

Marzari, F., Scholl, H., 1998b. The growth of Jupiter and Saturn and the capture of Trojans. Astron. Astroph. 339, 278-285 \\

Marzari, F., Scholl, H., Murray, C., Lagerkvist, C., 2002. Origin and Evolution of Trojan Asteroids. In Asteroids III, W. F. Bottke Jr., A. Cellino, P. Paolicchi, and R. P. Binzel (eds), University of Arizona Press, Tucson, 725-738.\\

Marzari, F., Tricarico, P., Scholl, H., 2003. Stability of Jupiter Trojans investigated using frequency map analysis: the MATROS project. MNRAS 345, 1091-1100.  \\

Milani, A., 1993. The Trojan asteroid belt: Proper elements, stability, chaos and families. Celest. Mech. Dynam. Astron. 57, 59-94. \\

Morbidelli, A., Levison, H. F., Tsiganis, K., Gomes, R., 2005. Chaotic capture of Jupiter's Trojan asteroids in the early Solar System. Nature  435, 462-465. \\

Moore, M.H., Donn, B., Khanna, R., A'Hearn, M.F., 1983. Studies of proton-irradiated cometary-type ice mixtures. Icarus 54, 388-405.\\
                                                                                
Moroz L., Baratta G., Strazzulla G., Starukhina L., Dotto E., Barucci M.A.,
Arnold G., Distefano E. 2004. Optical alteration of complex organics induced by ion irradiation: 1. Laboratory experiments suggest unusual space weathering trend. Icarus 170, 214-228. \\

Peixinho, N., Doressoundiram, A., Delsanti, A., Boehnhardt, H., Barucci, M. A., Belskaya, I., 2003. Reopening the TNOs color controversy: Centaurs bimodality and TNOs unimodality. Astron. Astrophys. 410, 29--32. \\

Shoemaker, E. M., Shoemaker, C. S., Wolfe, R. F., 1989. Trojan asteroids: populations, dynamical structure and origin of the L4 and L5 swarms. In Binzel, Gehrels, 
Matthews (Eds.), Asteroids II. Univ. of Arizona Press, Tucson, pp. 487-523. \\

Strazzulla, G., 1998. Chemistry of Ice Induced by Bombardment with Energetic
Charged Particles. In Solar System Ices (B. Schmitt, C. de Bergh, M. Festou,
eds.), Kluwer Academic, Dordrecht, Astrophys. Space Sci Lib. 281.\\

Thompson, W.R., Murray, B.G.J.P.T., Khare, B.N., Sagan, C. 1987. Coloration
and darkening of methane clathrate and other ices by charged particle
irradiation - Applications to the outer solar system. JGR 92, 14933-14947. \\
                                                                                
Xu, S., Binzel, R. P., Burbine, T. H., Bus, S. J., 1995. Small main-belt asteroid spectroscopic survey: Initial results. Icarus 115, 1--35. \\

Vilas, F. 1994.  A quick look method of detecting water of hydration in small 
solar system bodies. LPI 25, 1439-1440.\\

Zappala, V., Cellino, A., Farinella, P., Kne\u{z}evi\'{c}, Z., 1990. Asteroid 
families. I - Identification by hierarchical clustering and reliability assessment. 
Astron. J. 100,  2030-2046. \\

\newpage

{\bf Tables}

\begin{table}
\caption{Observing conditions of the investigated L5 asteroids. For each object 
we report the observational date and universal time, total exposure time, number of acquisitions with exposure time of each acquisition, airmass, and the observed solar analogs with their 
airmass.}
\label{spec_obsL5}
\begin{center}
\begin{tabular}{|l|c|c|c|c|c|c|} \hline
{\bf  Obj} & {\bf Date} &
{\bf UT} & {\bf T\boldmath{$_{exp}$ (s)}} & {\bf n\boldmath{$_{exp}$}} & {\bf air.} & {\bf
Solar An. (air.)} \\
\hline
{\bf Anchises} &&&&&& \\  \hline           
1173   & 17 Jan 05 & 06:06 &   60 & 1$\times$60s   & 1.42 & HD76151 (1.48) \\ 
23549  & 17 Jan 05 & 07:20 &  480 & 2$\times$240s  & 1.60 & HD76151 (1.48) \\
24452  & 17 Jan 05 & 07:54 &  960 & 4$\times$240s  & 1.44 & HD76151 (1.48) \\
47967  & 17 Jan 05 & 05:34 &  800 & 2$\times$400s  & 1.38 & HD76151 (1.48) \\   
2001 SB173&17 Jan 05&06:28 & 1200 & 2$\times$600s  & 1.35 & HD76151 (1.48) \\ 
\hline
{\bf Cloanthus}  &&&&&& \\  \hline   
5511   & 19 Jan 05 & 06:04 &  960 & 4$\times$240s  & 1.26 & HD76151 (1.12) \\
51359  & 19 Jan 05 & 04:13 &  660 & 1$\times$660s  & 1.36 & HD76151 (1.12) \\ 
\hline
{\bf Misenus}   &&&&&& \\  \hline    
11663  & 17 Jan 05 & 05:13 &  400 & 1$\times$400s  & 1.21 & HD44594 (1.12) \\ 
32794  & 18 Jan 05 & 03:13 & 1800 & 2$\times$900s  & 1.39 & HD28099 (1.44) \\
56968  & 17 Jan 05 & 04:31 &  400 & 2$\times$400s  & 1.21 & HD44594 (1.12) \\ 
1988 RE12 &18 Jan 05&04:12 & 2000 & 2$\times$1000s & 1.31 & HD28099 (1.44) \\
2000 SC51 &18 Jan 05&06:09 & 1320 & 2$\times$660s  & 1.16 & HD44594 (1.17) \\
2001 UY123&18 Jan 05&06:46 & 1320 & 2$\times$660s  & 1.32 & HD44594 (1.17) \\ 
\hline
{\bf Phereclos}  &&&&&& \\  \hline    
9030   & 18 Jan 05 & 08:19 & 1000 & 1$\times$1000s & 1.37 & HD44594 (1.17) \\
11488  & 19 Jan 05 & 03:31 & 1320 & 2$\times$660s  & 1.99 & HD76151 (1.12) \\
31820  & 19 Jan 05 & 07:02 & 1320 & 2$\times$660s  & 1.35 & HD76151 (1.11) \\   \hline
{\bf Sarpedon}  &&&&&& \\  \hline   
48252  & 18 Jan 05 & 02:32 & 1320 & 2$\times$660s  & 1.30 & HD28099 (1.44) \\  
84709  & 19 Jan 05 & 05:35 & 1320 & 2$\times$660s  & 1.34 & HD76151 (1.12) \\ 
\hline
{\bf Panthoos}  &&&&&& \\  \hline    
4829   & 17 Jan 05 & 08:37 &  720 & 3$\times$240s  & 1.45 & HD76151 (1.48) \\ 
30698  & 18 Jan 05 & 01:54 & 1320 & 2$\times$660s  & 1.73 & HD28099 (1.44) \\  
31821  & 18 Jan 05 & 05:27 & 1320 & 2$\times$660s  & 1.35 & HD28099 (1.44) \\
76804  & 17 Jan 05 & 03:35 & 1800 & 3$\times$600s  & 1.38 & HD44594 (1.12) \\
2001 VK85& 18 Jan 05&07:31 & 2000 & 2$\times$1000s & 1.23 & HD44594 (1.17) \\   
\hline
\end{tabular}
\end{center}
\end{table}

\begin{table}
\caption{Observing conditions of the investigated L4 asteroids. For each object 
we report the observational date and universal time, total exposure time, 
number of acquisitions with exposure time of each acquisition, airmass, and the observed solar analogs with their 
airmass.}
\label{spec_obsL4}
\begin{center}
\begin{tabular}{|l|c|c|c|c|c|c|} \hline
{\bf  Obj} & {\bf Date} &
{\bf UT} & {\bf T\boldmath{$_{exp}$ (s)}} & {\bf n\boldmath{$_{exp}$}} & {\bf air.} & {\bf
Solar An. (air.)} \\ \hline    
{\bf Eurybates} & & & && \\  \hline           
3548  & 25 May 04 &05:14 &600 & 2$\times$300s &  1.02 & SA107-684 (1.19)  \\
9818  & 26 May 04 &00:13 &780 & 1$\times$780s &  1.19 & SA102-1081(1.15)  \\
13862 & 25 May 04 &03:35 &1200& 2$\times$600s &  1.09 & SA107-998 (1.15)  \\
18060 & 25 May 04 &02:47 &1500& 2$\times$750s &  1.07 & SA107-998 (1.15)  \\
24380 & 25 May 04 &06:53 & 780& 1$\times$780s &  1.18 & SA107-684 (1.19)  \\
24420 & 25 May 04 &08:49 & 900& 1$\times$900s &  1.59 & SA112-1333 (1.17) \\
24426 & 26 May 04 &00:13 &1440& 2$\times$720s &  1.13 & SA107-684 (1.17)  \\
28958 & 26 May 04 &07:14 &1800& 2$\times$900s &  1.35 & SA107-684 (1.17)  \\
39285 & 25 May 04 &05:40 &2700& 3$\times$900s &  1.09 & SA107-684 (1.19)  \\
43212 & 25 May 04 &07:39 &2340& 3$\times$780s &  1.39 & SA110-361 (1.15)  \\
53469 & 25 May 04 &02:05 &1800& 2$\times$900s &  1.04 & SA107-998 (1.15)  \\
65150 & 26 May 04 &01:59 &3600& 4$\times$900s &  1.07 & SA102-1081 (1.20) \\
65225 & 26 May 04 &03:40 &3600& 4$\times$900s &  1.04 & SA107-684 (1.17)  \\
1996RD29 &26 May 04 &05:12&2700&3$\times$900s &  1.10 & SA107-684 (1.17)  \\
2000AT44 &25 May 04 &04:14&1800&2$\times$900s &  1.04 & SA107-684 (1.19)  \\ 
2002CT22 &26 May 04 &00:49&2400&4$\times$600s &  1.08 & SA102-1081 (1.15) \\
2002EN68 &26 May 04 &08:10&1800&2$\times$900s &  1.62 & SA107-684 (1.17)  \\
\hline
{\bf 1986~WD } & & & && \\  
\hline  
4035  & 10 Apr 03 & 03:28 & 600&1$\times$600s & 1.09 &  SA107-684 (1.15) \\
6545  & 10 Apr 03 & 02:39 & 900&1$\times$900s & 1.16 &  SA107-684 (1.15) \\
11351 & 10 Apr 03 & 09:21 & 900&1$\times$900s & 1.28 &  SA107-684 (1.15) \\
14707 & 11 Apr 03 & 08:11 &1200&1$\times$1200s& 1.15 &  SA107-684 (1.15) \\
24233 & 11 Apr 03 & 02:29 &1200&1$\times$1200s& 1.39 &  SA107-684 (1.37) \\
24341 & 11 Apr 03 & 05:47 & 900&1$\times$900s & 1.16 &  SA107-684 (1.17) \\
\hline
{\bf 1986~TS6} & & & && \\  \hline
12921 & 10 Apr 03 & 07:33 & 900 &1$\times$900s & 1.39 & SA107-684 (1.40) \\ 
\hline
\end{tabular}
\end{center}
\end{table}

\begin{table}
       \begin{center}
       \caption{Visible photometric observations of L4 and L5 Trojans (ESO-NTT EMMI): for each object, date, computed 
V magnitude, B-V, V-R and V-I colors are reported. The given UT is for the V filter acquisition. The observing photometric sequence (V-R-B-I) took a few minutes.}
        \label{photometry}
\footnotesize{
\begin{tabular}{|l|l|c|c|c|c|c|} \hline
Object  &  date & UT & V & B-V & V-R & V-I \\ \hline \hline
{\bf L4} & & & & & & \\ \hline \hline
{\bf 1986~WD}  & & &&&&\\  \hline
4035  & 10 Apr 03& 03:11& 16.892$\pm$0.031 &  0.752$\pm$0.040&   0.473$\pm$0.042&  0.926$\pm$0.055 \\
4035  & 10 Apr 03& 04:22& 16.981$\pm$0.031 &  0.752$\pm$0.040&   0.495$\pm$0.042&  0.945$\pm$0.055  \\
6545  & 10 Apr 03& 02:22& 17.558$\pm$0.031 &  0.734$\pm$0.041&   0.499$\pm$0.042&  0.935$\pm$0.055 \\ 
11351 & 10 Apr 03& 09:03& 18.407$\pm$0.032 &  0.739$\pm$0.044&   0.498$\pm$0.044&  0.900$\pm$0.057  \\
14707 & 11 Apr 03& 06:46& 18.666$\pm$0.031 &  0.751$\pm$0.041  &0.401$\pm$0.033  & 0.804$\pm$0.055  \\
14707 & 11 Apr 03& 08:37& 18.873$\pm$0.031 &  0.754$\pm$0.041  &0.424$\pm$0.033  & 0.790$\pm$0.056  \\
24233 & 11 Apr 03& 01:33& 18.894$\pm$0.034 &  0.704$\pm$0.051  &0.481$\pm$0.037  & 0.899$\pm$0.058  \\
24341 & 11 Apr 03& 05:05& 19.376$\pm$0.032 &  0.713$\pm$0.043  &0.369$\pm$0.035  & 0.759$\pm$0.057  \\ \hline
{\bf 1986~TS6 }  &&&&&&\\ \hline
12921 & 10 Apr 03& 07:12& 18.393$\pm$0.031 &  0.673$\pm$0.040&   0.421$\pm$0.042&  0.786$\pm$0.055  \\ \hline \hline
{\bf L5} &cut off &150m/s & & & & \\ \hline  \hline
{\bf Anchises}  & & & & & & \\   \hline
1173  & 17 Jan 05 & 05:54 & 16.595$\pm$0.024 & 0.811$\pm$0.034 & 0.402$\pm$0.035 & 0.805$\pm$0.038 \\ 
23549 & 17 Jan 05 & 07:09 & 18.969$\pm$0.050 & 0.800$\pm$0.071 & 0.485$\pm$0.068 & 0.872$\pm$0.075 \\
24452 & 17 Jan 05 & 07:48 & 18.757$\pm$0.043 & 0.872$\pm$0.056 & 0.441$\pm$0.056 & 0.847$\pm$0.066 \\
47967 & 17 Jan 05 & 05:27 & 19.382$\pm$0.044 & 0.899$\pm$0.058 & 0.489$\pm$0.069 & 0.965$\pm$0.075   \\
2001 SB173 & 17 Jan 05 & 06:20 & 19.882$\pm$0.043 & 0.992$\pm$0.060 & 0.503$\pm$0.064 & 0.927$\pm$0.078 \\ \hline
{\bf Cloanthus} & & & & & & \\  \hline
5511 & 19 Jan 05 & 05:52 &      17.968$\pm$0.020 &   0.906$\pm$0.027 &  0.442$\pm$0.027  &  0.968$\pm$0.032 \\
51359  & 19 Jan 05 & 03:54 &    19.631$\pm$0.102 &   0.864$\pm$0.201 &  0.447$\pm$0.131  &  0.885$\pm$0.164  \\ \hline
{\bf Misenus}  & & & & & & \\  \hline
11663 & 17 Jan 05 & 05:05 & 18.473$\pm$0.022 & 0.837$\pm$0.030 & 0.409$\pm$0.030 & 0.872$\pm$0.039 \\
32794  & 18 Jan 05 & 03:07 & 19.685$\pm$0.038 &   0.923$\pm$0.065 &  0.393$\pm$0.056  &  0.879$\pm$0.057 \\ 
56968 & 17 Jan 05 & 04:18 & 18.596$\pm$0.026 & 0.986$\pm$0.040 & 0.494$\pm$0.033 & 1.003$\pm$0.036   \\
1988 RE12  & 18 Jan 05 & 04:00 & 20.892$\pm$0.081 &   0.826$\pm$0.132 &  0.388$\pm$0.108  &  0.871$\pm$0.106  \\
2000 SC51  & 18 Jan 05 & 06:03 & 19.876$\pm$0.038 &   1.016$\pm$0.055 &  0.444$\pm$0.059  &  0.896$\pm$0.056  \\
2001 UY123 & 18 Jan 05 & 06:41 & 19.869$\pm$0.047 &   0.890$\pm$0.058 &  0.537$\pm$0.056  &  0.971$\pm$0.063  \\ \hline
{\bf Phereclos} & & & & & & \\  \hline
9030   & 18 Jan 05 & 08:14 &    18.397$\pm$0.020 &   0.887$\pm$0.024 &  0.493$\pm$0.027  &  0.973$\pm$0.028 \\
11488  & 19 Jan 05 & 02:57 &    18.931$\pm$0.066 &   0.868$\pm$0.101 &  0.430$\pm$0.079  &  0.848$\pm$0.084 \\
31820  & 19 Jan 05 & 06:39 &    20.041$\pm$0.077 &   0.889$\pm$0.093 &  0.520$\pm$0.091  &  0.916$\pm$0.123 \\ \hline
{\bf Sarpedon} & & & & & & \\  \hline
48252  & 18 Jan 05 & 02:25 &    19.878$\pm$0.060 &   0.949$\pm$0.100 &  0.467$\pm$0.093 &   0.903$\pm$0.090 \\
84709  & 19 Jan 05 & 05:10 &    19.862$\pm$0.068 &   0.855$\pm$0.087 &  0.462$\pm$0.090  &  1.010$\pm$0.094  \\ \hline
{\bf Panthoos} & & & & & & \\  \hline
4829  & 17 Jan 05 & 08:18 & 18.430$\pm$0.029 & 0.851$\pm$0.050 & 0.420$\pm$0.039 & 0.792$\pm$0.052 \\
30698  & 18 Jan 05 & 01:45 &    19.353$\pm$0.036 &    --             &  0.472$\pm$0.042  &  0.865$\pm$0.047 \\
31821  & 18 Jan 05 & 05:21 &    19.328$\pm$0.076 &   0.980$\pm$0.111 &  0.440$\pm$0.097  &  0.901$\pm$0.108 \\ 
76804 & 17 Jan 05 & 03:21 & 19.471$\pm$0.065  &  0.803$\pm$0.082 & 0.446$\pm$0.070 &   0.889$\pm$ 0.080  \\
2001 VK85  & 18 Jan 05 & 07:23 & 20.179$\pm$0.038 &   0.822$\pm$0.063 &  0.462$\pm$0.048  &  1.020$\pm$0.050  \\ 
\hline
\end{tabular}
}
\end{center}
\end{table}

\begin{table}
\caption{L5 families. We report for each target the absolute magnitude H 
and the estimated diameter (diameters marked by $\ast$ are taken from IRAS data), the spectral slope $S$ computed between 5500 and 
8000 \AA\, and the taxonomic class (T) derived following Dahlgren \& Lagerkvist 
(1995) classification scheme. The asteroids marked with $^{a}$ were observed by Fornasier et al. (2004a), and their spectral slope values have been recomputed in the 5500-8000 \AA\ wavelength range; asteroids 23694, 30698 and 32430, 
previously Astyanax members, have been reassigned to the Panthoos family due to refined proper elements.}
\label{spec_L5}
\begin{center}
\footnotesize{
\begin{tabular}{|l|c|c|c|c|} \hline
{\bf  Obj} & {\bf H} & {\bf D (km)} & {\bf S (\%/\boldmath{$10^{3}$\AA)}}  & {\bf T} \\
\hline                                               
{\bf Anchises} &&&& \\  \hline           
1173  &  8.99 & $^{\ast}$126$^{+11}_{-11}$   &  3.87$\pm$0.70 & P \\ 
23549 &  12.04 &         26$^{+4}_{-6}$    &  8.49$\pm$0.88 & D \\
24452 &  11.85 &         29$^{+5}_{-7}$    &  7.42$\pm$0.70 & D \\
47967 &  12.15 &          25$^{+4}_{-6}$    &  9.21$\pm$0.78 & D \\   
2001 SB173 & 12.77     & 19$^{+3}_{-5}$    & 14.78$\pm$0.99 & D \\ 
\hline
{\bf Cloanthus}  &&&& \\  \hline   
5511  &    10.43  & 55$^{+8}_{-13}$   & 10.84$\pm$0.65 & D \\
51359 &    12.25  & 24$^{+6}_{-4}$    & 12.63$\pm$1.30 & D \\ 
\hline
{\bf Misenus}  & &&& \\  \hline    
11663 &    10.95     &   44$^{+7}_{-10}$   &  6.91$\pm$0.70 & DP \\ 
32794 &    12.77     &   19$^{+3}_{-5}$    &  6.59$\pm$0.88 & DP \\     
56968 &    11.72     &   30$^{+5}_{-7}$    & 15.86$\pm$0.71 & D  \\ 
1988 RE12  & 13.20   &   16$^{+2}_{-4}$    &  4.68$\pm$1.20 & P  \\      
2000 SC51  &  12.69  &   20$^{+3}_{-5}$    &  6.54$\pm$0.98 & DP \\
2001 UY123 &  12.75  &   19$^{+3}_{-5}$    &  8.28$\pm$0.88 & D  \\ 
\hline
{\bf Phereclos}  &&&& \\  \hline   
$^{a}$2357 & 8.86 & $^{\ast}$95$^{+4}_{-4}$ &  9.91$\pm$0.68 & D  \\
$^{a}$6998 & 11.43 & 34$^{+5}_{-8}$ & 11.30$\pm$0.75 & D  \\
9030  &  11.14   & 40$^{+6}_{-10}$ & 10.35$\pm$0.76 & D  \\
$^{a}$9430 &  11.47  & 35$^{+5}_{-8}$ & 10.02$\pm$0.90 & D  \\
11488  &    11.82   &  29$^{+5}_{-7}$ &  5.37$\pm$0.92 & PD \\
$^{a}$18940  & 11.81 & 29$^{+4}_{-7}$ &  7.13$\pm$0.75 & D  \\
31820  &      12.63       & 20$^{+3}_{-5}$ &  7.53$\pm$0.80 & D  \\   
\hline
{\bf Sarpedon}  &&&& \\  \hline   
$^{a}$2223 & 9.25  & $^{\ast}$95$^{+4}_{-4}$ & 10.20$\pm$0.65 & D  \\
$^{a}$5130 & 9.85   &   71$^{+11}_{-18}$    & 10.45$\pm$0.65 & D  \\
$^{a}$17416  &  12.83 &18$^{+3}_{-5}$      & 10.80$\pm$0.90 & D  \\
$^{a}$25347  &  11.59 &33$^{+5}_{-8}$      & 10.11$\pm$0.83 & D  \\
48252  &   12.84      &18$^{+3}_{-5}$      &  9.62$\pm$0.82 & D  \\     
84709  &   12.70      &19$^{+3}_{-5}$      & 11.64$\pm$0.84 & D  \\ 
\hline
{\bf Panthoos}  &&&& \\  \hline    
4829   &   11.16      &39$^{+6}_{-10}$     &  5.03$\pm$0.70 & PD \\ 
$^{a}$23694  & 11.61  & 32$^{+5}_{-8}$      &  8.20$\pm$0.72 & D  \\
30698  &    12.14     &25$^{+4}_{-6}$      &  8.23$\pm$1.00 & D  \\ 
$^{a}$30698 &  12.27  &25$^{+4}_{-6}$      &  9.08$\pm$0.82 & D  \\ 
$^{a}$32430 &  12.23  & 25$^{+4}_{-6}$      &  8.12$\pm$1.00 & D  \\
31821  &   11.99      &27$^{+4}_{-6}$      & 10.58$\pm$0.82 & D  \\       
76804  &   12.16      &25$^{+4}_{-6}$      &  7.29$\pm$0.71 & D  \\
2001 VK85 & 12.79     &19$^{+3}_{-5}$      & 14.39$\pm$0.81 & D  \\   
\hline
\end{tabular}
}
\end{center}
\end{table}

\begin{table}
\small{
\caption{L4 Families. We report for each target the absolute magnitude H and the estimated diameter (diameters marked by $\ast$ are taken from IRAS data, while absolute magnitudes marked by $\ast\ast$ are taken from the astorb.dat file of the Lowell Observatory), the spectral slope $S$ computed between 5500 and 8000 \AA, and the taxonomic class (T) derived following Dahlgren \& Lagerkvist 
(1995) classification scheme. 
The asteroids marked with $^{a}$ were observed by Dotto et al. (2006), and their 
spectral slope values have been recomputed in the 5500-8000 \AA\ wavelength range.}
\label{spec_L4}
\begin{center}
\begin{tabular}{|l|c|c|c|c|} \hline
{\bf  Obj} & {\bf H} & {\bf D (km)} & {\bf S (\%/\boldmath{$10^{3}$\AA)}}  & {\bf T} \\
\hline 
{\bf Eurybates} &&&& \\  \hline           
3548        & 9.50$^{\ast\ast}$ & $^{\ast}$72$^{+4}_{-4}$  &-0.18$\pm$0.57  & C \\
9818        & 11.00$^{\ast\ast}$ & 42$^{+6}_{-10}$ & 2.12$\pm$0.72  & P \\ 
13862       & 11.10$^{\ast\ast}$ & 40$^{+6}_{-10}$ & 1.59$\pm$0.70  & C \\
18060       & 11.10$^{\ast\ast}$ & 40$^{+6}_{-10}$ & 2.86$\pm$0.60  & P \\   
24380       & 11.20$^{\ast\ast}$ & 38$^{+6}_{-9}$  & 0.34$\pm$0.65  & C \\   
24420       & 11.50$^{\ast\ast}$  & 33$^{+5}_{-8}$  & 1.65$\pm$0.70  & C \\ 
24426       & 12.50$^{\ast\ast}$ & 21$^{+3}_{-5}$  & 4.64$\pm$0.80  & P \\
28958       & 12.10$^{\ast\ast}$  & 25$^{+4}_{-6}$  &-0.04$\pm$0.80  & C \\
39285       & 12.90$^{\ast\ast}$ & 17$^{+3}_{-4}$  & 0.25$\pm$0.69  & C \\
43212       & 12.30$^{\ast\ast}$ & 23$^{+4}_{-6}$  & 1.19$\pm$0.78  & C \\
53469       & 11.80$^{\ast\ast}$ & 29$^{+4}_{-7}$  & 0.17$\pm$0.80  & C \\
65150       & 12.90$^{\ast\ast}$  & 17$^{+3}_{-4}$  & 4.14$\pm$0.70  & P \\  
65225       & 12.80$^{\ast\ast}$  & 18$^{+3}_{-4}$  & 0.97$\pm$0.85  & C \\
1996RD29    & 13.06$^{\ast\ast}$  & 16$^{+3}_{-4}$  & 2.76$\pm$0.89  & P \\
2000AT44    & 12.16$^{\ast\ast}$  & 24$^{+3}_{-6}$  &-0.53$\pm$0.83  & C \\ 
2002CT22    & 12.04$^{\ast\ast}$  & 26$^{+4}_{-6}$  & 2.76$\pm$0.73  & P \\
2002EN68    & 12.30$^{\ast\ast}$  & 23$^{+3}_{-6}$  & 3.60$\pm$0.98  & P \\
\hline         
{\bf 1986~WD} &&&& \\  \hline
4035        & 9.72  & $^{\ast}$68$^{+5}_{-5}$  &  9.78$\pm$0.61     & D \\       
$^a$4035     & 9.30$^{\ast\ast}$ & $^{\ast}$68$^{+5}_{-5}$ & 15.19$\pm$0.61     & D \\
6545         & 10.42  & 55$^{+8}_{-13}$ & 11.32$\pm$0.63     & D \\
$^a$6545     & 10.00$^{\ast\ast}$ &   66$^{+10}_{-16}$  &  9.88$\pm$0.56     & D \\
11351        & 10.88 & 44$^{+7 }_{-11}$ & 10.26$\pm$0.67     & D \\ 
$^a$11351    & 10.50$^{\ast\ast}$ & 53$^{+8 }_{-13}$ & 10.44$\pm$0.61     & D \\ 
14707        & 11.25 &38$^{+6}_{-9.4}$ & -1.06$\pm$1.00     & C \\       
24233        & 11.58 &33$^{+5}_{-8.0}$ &  6.37$\pm$0.67    &DP \\      
24341        & 11.99 & 27$^{+4}_{-6}$   & -0.26$\pm$0.71    & C \\
\hline         
{\bf 1986~TS6} & &&& \\  \hline   
12917       & 11.61 & 32$^{+5}_{-8}$  &10.98$\pm$0.68 &  D \\           
%12917       & 32$^{+5}_{-8}$  &10.743$\pm$0.150 &  D \\          
12921       & 11.12 & 40$^{+6}_{-10}$ & 4.63$\pm$0.75 &  P \\     
$^a$12921   & 10.70$^{\ast\ast}$ & 48$^{+7}_{-12}$ & 3.74$\pm$1.00 &  P \\
13463       & 11.27 & 37$^{+6}_{-9} $ & 4.37$\pm$0.65 &  P \\    
%13463       & 37$^{+6}_{-9} $ & 5.123$\pm$0.116 &  PD \\   
15535       & 10.70 & 48$^{+7}_{-12}$ &10.67$\pm$0.65  & D \\            
%15535       & 48$^{+7}_{-12}$ &11.354$\pm$0.131 & D \\            
20738       & 11.67 & 31$^{+5}_{-8} $ & 8.84$\pm$0.70  & D \\           
%20738       & 31$^{+5}_{-8} $ & 9.190$\pm$0.169 & D \\           
24390       & 11.80 & 29$^{+5}_{-7} $ & 9.53$\pm$0.62  & D \\  
\hline
\end{tabular}
\end{center}
}
\end{table}

\begin{sidewaystable}
\caption{Results of the statistical analysis on the spectral slope distribution as a function of the diameters. For
each test bin, the average slope and the dispersion are listed; the
size of the sample is reported in parenthesis.  For each pair of
subsamples, the probability that both are randomly extracted from the
same global sample is listed, as estimated by the t-, f- and ks-test,
respectively. Low probability indicates significant differences between the
subsamples. }
\label{tab:statSlopes}
\begin{center}
\begin{tabular}{|l|ccccc|} 
\hline  
 Diameter range       &  0--25 km            &    25--50 km         &  50--75 km                    &      75--100 km          &     $>$100 km        \\   \hline\hline
  $S$ average$\pm\sigma$ &7.17$\pm$4.79 (22)&6.92$\pm$4.69 (48)&8.91$\pm$4.68 (26)  & 6.74$\pm$5.85 (21)& 7.87$\pm$2.88 (21)   \\  
(\%/10$^{3}$\AA) & & & & &  \\ \hline
  0--25                &                &  0.842 0.876 0.579 &  0.213 0.903 0.575 &  0.792 0.370 0.775&  0.551 0.017 0.494    \\
 25--50                &                &                    & 0.088 0.985 0.150  & 0.897 0.216 0.519 & 0.286 0.011 0.275    \\
 50--75                &                &                    &                    & 0.176 0.289 0.469 & 0.344 0.019 0.440    \\
 75--100               &                &                    &                    &                   & 0.442 0.001 0.469    \\
\hline
\end{tabular}

\end{center}
\end{sidewaystable}

\begin{sidewaystable}
\caption{Mean color indices and spectral slope of different classes of minor bodies of the outer Solar System. For each class the number of objects considered is also listed.}
\label{tab:average}
\begin{center}
\begin{tabular}{lrrrrrrrr}
\hline
Color& Plutinos& Cubewanos& Centaurs& Scattered& Comets & Trojans \\
\hline
\hline
B-V &  36&  87&  29&  33&   2&  74\\
   &   0.895$\pm$  0.190&   0.973$\pm$  0.174&   0.886$\pm$  0.213&   0.875$\pm$  0.159&   0.795$\pm$  0.035&   0.777$\pm$  0.091\\
V-R &  38&  96&  30&  34&  19&  80\\
   &   0.568$\pm$  0.106&   0.622$\pm$  0.126&   0.573$\pm$  0.127&   0.553$\pm$  0.132&   0.441$\pm$  0.122&   0.445$\pm$  0.048\\
V-I &  34&  64&  25&  25&   7&  80\\
   &   1.095$\pm$  0.201&   1.181$\pm$  0.237&   1.104$\pm$  0.245&   1.070$\pm$  0.220&   0.935$\pm$  0.141&   0.861$\pm$  0.090\\
V-J &  10&  14&  11&   8&   1&  12\\
   &   2.151$\pm$  0.302&   1.750$\pm$  0.456&   1.904$\pm$  0.480&   2.041$\pm$  0.391&   1.630$\pm$  0.000&   1.551$\pm$  0.120\\
V-H &   3&   7&  11&   4&   1&  12\\
   &   2.698$\pm$  0.083&   2.173$\pm$  0.796&   2.388$\pm$  0.439&   2.605$\pm$  0.335&   1.990$\pm$  0.000&   1.986$\pm$  0.177\\
V-K &   2&   5&   9&   2&   1&  12\\
   &   2.763$\pm$  0.000&   2.204$\pm$  1.020&   2.412$\pm$  0.396&   2.730$\pm$  0.099&   2.130$\pm$  0.000&   2.125$\pm$  0.206\\
\\
R-I &  34&  64&  25&  26&   8&  80\\
   &   0.536$\pm$  0.135&   0.586$\pm$  0.148&   0.548$\pm$  0.150&   0.517$\pm$  0.102&   0.451$\pm$  0.059&   0.416$\pm$  0.057\\
J-H &  11&  17&  21&  11&   1&  12\\
   &   0.403$\pm$  0.292&   0.370$\pm$  0.297&   0.396$\pm$  0.112&   0.348$\pm$  0.127&   0.360$\pm$  0.000&   0.434$\pm$  0.064\\
H-K &  10&  16&  20&  10&   1&  12\\
   &  -0.034$\pm$  0.171&   0.084$\pm$  0.231&   0.090$\pm$  0.142&   0.091$\pm$  0.136&   0.140$\pm$  0.000&   0.139$\pm$  0.041\\
\\

Slope &  38&  91&  30&  34&   8&  80\\
(\%/10$^{3}$\AA)   &  19.852$\pm$ 10.944&  25.603$\pm$ 13.234&  20.601$\pm$ 13.323&  18.365$\pm$ 12.141&  10.722$\pm$  6.634&   7.241$\pm$  3.909\\
\hline
\end{tabular}
\end{center}
\end{sidewaystable}

\begin{sidewaystable}
\caption{Statistical tests performed to compare the color and slope distributions of different classes of minor bodies (Plt= Plutinos, Resonant TNOs; QB1=
  Cubiwanos, Classical TNOs; Cent= Centaurs; Scat= scattered TNOs; Com=
  Short Period Comet nuclei) with those of Trojans. The first five columns
  consider all the Trojans, the second five only the Eurybates family, the
  third five only the non-Eurybates family Trojans. For each color, the first line shows the number of objects used for the comparison (2nd is the number of Trojans), and the second line reports the
  probability resulting from the test. A very low value indicates that the two compared distributions are {\it not} statistically 
  compatible. Probabilities are in boldface when the size of the
  samples is large enough for the value to be meaningful.}
\label{tab:tests}
\begin{center}
\tiny
\begin{tabular}{l|rrrrr|rrrrr|rrrrr}
\hline											    
\multicolumn{10}{l}{\bf  f-test }\\
\hline 
Color& 
\multicolumn{5}{c|}{All Trojans}&
\multicolumn{5}{c|}{Only Eurybates}&
\multicolumn{5}{c}{Only NON-Eurybates}
\\
& Plt     & QB1    & Cent   & Scat   &Com                           & Plt     & QB1   & Cent& Scat &Com                                & Plt & QB1 & Cent & Scat & Com \\ 
\hline											    
\hline										    
B-V	&  36  74&  83  74&  29  74&  33  74&   2  74						    &  36  14&  83  14&  29  14&  33  14&   2  14                                          &  36  60&  83  60&  29  60&  33  60&   2  60\\                          
  	&\bf     0.000&\bf     0.000&\bf     0.000&\bf     0.000&     0.600	    &\bf     0.001&\bf     0.001&\bf     0.000&\bf     0.005&     0.722     &\bf     0.000&\bf     0.000&\bf     0.000&\bf     0.000&     0.598\\   
V-R	&  38  80&  92  80&  30  80&  34  80&  19  80						    &  38  17&  92  17&  30  17&  34  17&  19  17                                          &  38  63&  92  63&  30  63&  34  63&  19  63\\                            
  	&\bf     0.000&\bf     0.000&\bf     0.000&\bf     0.000&\bf     0.000	    &\bf    0.000&\bf    0.000&\bf    0.000&\bf    0.000&\bf    0.000  &\bf     0.000&\bf    0.000&\bf    0.000&\bf    0.000&\bf     0.000\\
R-I	&  34  80&  62  80&  25  80&  26  80&   8  80						    &  34  17&  62  17&  25  17&  26  17&   8  17                                          &  34  63&  62  63&  25  63&  26  63&   8  63\\                       
  	&\bf     0.000&\bf     0.000&\bf     0.000&\bf     0.000&\bf     0.773	    &\bf     0.000&\bf    0.000&\bf    0.000&\bf     0.001&\bf     0.335  &\bf     0.000&\bf    0.000&\bf    0.000&\bf     0.000&\bf     0.185\\
Slope	&  38  80&  87  80&  30  80&  34  80&   8  80						    &  38  17&  87  17&  30  17&  34  17&   8  17                                          &  38  63&  87  63&  30  63&  34  63&   8  63\\                             
  	&\bf     0.000&\bf    0.000&\bf    0.000&\bf     0.000&\bf     0.020	    &\bf    0.000&\bf    0.000&\bf    0.000&\bf    0.000&\bf     0.000  &\bf    0.000&\bf    0.000&\bf    0.000&\bf    0.000&\bf     0.000\\
\hline
\hline
\multicolumn{10}{l}{\bf  t-test }\\
\hline 
Color& 
\multicolumn{5}{c|}{All Trojans}&
\multicolumn{5}{c|}{Only Eurybates}&
\multicolumn{5}{c}{Only NON-Eurybates}
\\
& Plt     & QB1    & Cent   & Scat   &Com                           & Plt     & QB1   & Cent& Scat &Com                                & Plt & QB1 & Cent & Scat & Com \\ 
\hline
\hline
B-V &  36  74&  83  74&  29  74&  33  74&   2  74					       &  36  14&  83  14&  29  14&  33  14&   2  14                                            &  36  60&  83  60&  29  60&  33  60&   2  60\\                                       
   &\bf     0.001&\bf     0.000&\bf     0.012&\bf     0.002&     0.608	       &\bf     0.000&\bf     0.000&\bf     0.001&\bf     0.000&     0.139       &\bf     0.003&\bf     0.000&\bf     0.025&\bf     0.006&     0.858\\   
V-R &  38  80&  92  80&  30  80&  34  80&  19  80					       &  38  17&  92  17&  30  17&  34  17&  19  17                                            &  38  63&  92  63&  30  63&  34  63&  19  63\\                                       
   &\bf     0.000&\bf    0.000&\bf     0.000&\bf     0.000&\bf    0.916     &\bf     0.000&\bf    0.000&\bf     0.000&\bf     0.000&\bf     0.083    &\bf     0.000&\bf    0.000&\bf     0.000&\bf     0.000&\bf    0.532\\
R-I &  34  80&  62  80&  25  80&  26  80&   8  80					       &  34  17&  62  17&  25  17&  26  17&   8  17                                            &  34  63&  62  63&  25  63&  26  63&   8  63\\                                       
   &\bf     0.000&\bf     0.000&\bf     0.000&\bf     0.000&\bf     0.154     &\bf     0.000&\bf    0.000&\bf     0.000&\bf     0.000&\bf     0.001    &\bf     0.000&\bf     0.000&\bf     0.001&\bf     0.000&\bf     0.502\\
Slope &  38  80&  87  80&  30  80&  34  80&   8  80					       &  38  17&  87  17&  30  17&  34  17&   8  17                                            &  38  63&  87  63&  30  63&  34  63&   8  63\\                                       
   &\bf     0.000&\bf    0.000&\bf     0.000&\bf     0.000&\bf     0.185     &\bf     0.000&\bf    0.000&\bf     0.000&\bf     0.000&\bf     0.008    &\bf     0.000&\bf    0.000&\bf     0.000&\bf     0.000&\bf     0.404\\
\hline
\hline
\multicolumn{10}{l}{\bf  KS-test }\\
\hline 
Color& 
\multicolumn{5}{c|}{All Trojans}&
\multicolumn{5}{c|}{Only Eurybates}&
\multicolumn{5}{c}{Only NON-Eurybates}
\\
& Plt     & QB1    & Cent   & Scat   &Com                           & Plt     & QB1   & Cent& Scat &Com                                & Plt & QB1 & Cent & Scat & Com \\ 

\hline
\hline
B-V &  36  74&  83  74&  29  74&  33  74&   2  74					     &  36  14&  83  14&  29  14&  33  14&   2  14                                           &  36  60&  83  60&  29  60&  33  60&   2  60\\                                       
   &\bf     0.001&\bf     0.000&\bf     0.001&\bf     0.004&     0.330	     &\bf     0.002&\bf     0.000&\bf     0.035&\bf     0.000&     0.065      &\bf     0.003&\bf     0.000&\bf     0.002&\bf     0.047&     0.468\\   
V-R &  38  80&  92  80&  30  80&  34  80&  19  80					     &  38  17&  92  17&  30  17&  34  17&  19  17                                           &  38  63&  92  63&  30  63&  34  63&  19  63\\                                       
   &\bf     0.000&\bf     0.000&\bf     0.000&\bf     0.000&\bf     0.040     &\bf     0.000&\bf     0.000&\bf     0.000&\bf     0.000&\bf     0.008   &\bf     0.000&\bf     0.000&\bf     0.000&\bf     0.000&\bf     0.056\\
R-I &  34  80&  62  80&  25  80&  26  80&   8  80					     &  34  17&  62  17&  25  17&  26  17&   8  17                                           &  34  63&  62  63&  25  63&  26  63&   8  63\\                                       
   &\bf     0.000&\bf     0.000&\bf     0.000&\bf     0.000&\bf     0.201     &\bf     0.000&\bf     0.000&\bf     0.000&\bf     0.000&\bf     0.000   &\bf     0.000&\bf     0.000&\bf     0.000&\bf     0.000&\bf     0.587\\
Slope &  38  80&  87  80&  30  80&  34  80&   8  80					     &  38  17&  87  17&  30  17&  34  17&   8  17                                           &  38  63&  87  63&  30  63&  34  63&   8  63\\                                       
   &\bf     0.000&\bf     0.000&\bf     0.000&\bf     0.000&\bf     0.088     &\bf     0.000&\bf     0.000&\bf     0.000&\bf     0.000&\bf     0.002   &\bf     0.000&\bf     0.000&\bf     0.000&\bf     0.000&\bf     0.211\\
\hline
\end{tabular}
\end{center}
\end{sidewaystable}

%%%%%%%%%%%%%%%%%%%%%%%%%%%%%%%%%%%%%%%%%%%%%%%%%%%%%%%%%%%%%%%%%%%%%%%%%%%%%%
\newpage

{\bf Figure captions}
                                                                                             
\vspace{1cm}
                                                                                             
Fig. 1 - Reflectance spectra of 5 Anchises family members (L5 swarm). The photometric color indices are also converted to relative reflectance and overplotted on each spectrum. Spectra and photometry are shifted by 0.5 in reflectance for clarity. \\

Fig. 2 - Reflectance spectra of 6 Misenus family members (L5 swarm). The photometric color indices are also converted to relative reflectance and overplotted on each spectrum. Spectra and photometry are shifted by 0.5 in reflectance for clarity. \\

Fig. 3 - Reflectance spectra of 5 Panthoos family members (L5 swarm). The photometric color indices are also converted to relative reflectance and overplotted on each spectrum. Spectra and photometry are shifted by 0.5 in reflectance for clarity. For asteroid 30698, the B-V color is missing as a B filter
measurement was not available.  \\

Fig. 4 - Reflectance spectra of 2 Cloantus, 3 Phereclos and 2 Sarpedon family members (L5 swarm). The photometric color indices are also converted to relative reflectance and overplotted on each spectrum. Spectra and photometry are shifted by 1.0 in reflectance for clarity. \\

Fig. 5 - Reflectance spectra of the 17 Eurybates family members (L4 swarm). Spectra are shifted by 0.5 in reflectance for clarity. \\
 
Fig. 6 - Reflectance spectra of the 6 1986~WD family members and 12921, which is a member of the 1986~TS6 family (all belonging to the L4 swarm). Spectra are shifted by 1.0 in reflectance for clarity. \\

Fig. 7 - Plot of the spectral slope versus the estimated diameter for the families observed in the L5 swarm. \\

Fig. 8 - Plot of the spectral slope versus the estimated diameter for the families observed in the L4 swarm. \\

Fig. 9 - Plot of the observed spectral slopes versus the estimated diameter for the whole population of Jupiter Trojans investigated by us and available from the literature. The errors on slopes and diameters are not plotted to avoid confusion.  \\

Fig. 10 - Histogram of L5 Trojans taxonomic classes. \\

Fig. 11 - Histogram of L4 Trojans taxonomic classes ({\it Neg} indicates objects with negative spectral slope). \\

Fig. 12 - Color distributions as functions of the absolute magnitude
 	 $M(1,1)$, the inclination $i$ [degrees], the orbital semi-major axis $a$ [AU], the perihelion distance $q$ [AU], the
 	 eccentricity $e$, and the orbital energy
 	 $E$ (see text for definition). We include all the 
        available colors for distant minor bodies (TNOs, Centaurs, and cometary nuclei, see Hainaut \& Delsanti 2002). The Plutinos (resonant TNOs) are red filled triangles, Cubiwanos (classical TNOs) are pink filled circles, Centaurs are green open
 	 triangles, Scattered TNOs are blue open circles, and Trojans are cyan filled triangles. \\

Fig. 13 - $V-R$ versus $R-I$ color-color diagram for the observed Trojans and all distant minor bodies available in the updated Hainaut \& Delsanti
        (2002) database. The solid symbols are for the Trojans (square for Eurbybates,
        triangles for others). The open symbols are used as
        following: triangles for Plutinos, circles for Cubiwanos,
        squares for Centaurs, pentagons for Scattered, and starry
        square for Comets. The continuous line represents the "reddening line", that is the locus of objects with a linear reflectivity spectrum. 
The star symbol represents the Sun. \\

Fig. 14 - Cumulative function and histograms of the $B-V$ and $V-R$              color distributions and of the spectral slope for all the considered classes of objects. The dotted line marks the solar colors. \\

\end{document}